\documentclass[final,3p,times,twocolumn]{elsarticle}

\usepackage{amssymb}
\usepackage{amsmath}
\usepackage{enumitem}
\usepackage{placeins}
\usepackage[colorlinks=true,linkcolor=blue,citecolor=blue,urlcolor=blue]{hyperref}
\usepackage{subcaption}

\journal{Astroparticle Physics}

\begin{document}

\begin{frontmatter}

\title{Gamma rays as leptonic portals to energetic neutrinos: a new Monte Carlo approach}

\author[IFT,DFT]{Gaetano Di Marco}
\ead{gaetano.dimarco@ift.csic.es}
\author[IAP,LPNHE,IFT,DFT]{Rafael {Alves Batista}}
\author[IFT,DFT]{Miguel A. Sánchez-Conde}

\affiliation[IFT]{organization={Instituto de Física Teórica UAM/CSIC},
            addressline={Calle Nicolás Cabrera 13-15}, 
            city={Cantoblanco},
            postcode={28049 Madrid}, 
            country={Spain}}

\affiliation[DFT]{organization={Departamento de Física Teórica, M-15, Universidad Autónoma de Madrid},
            addressline={Calle Nicolás Cabrera 13-15}, 
            postcode={E-28049 Madrid}, 
            country={Spain}}

\affiliation[IAP]{organization={Institut d'Astrophysique de Paris (IAP), Sorbonne Université, CNRS, UMR 7095},
            addressline={98 bis bd Arago}, 
            postcode={75014 Paris}, 
            country={France}}

\affiliation[LPNHE]{Laboratoire de Physique Nucléaire et de Hautes Energies (LPNHE), Sorbonne Université, CNRS
            addressline={4 place Jussieu}, 
            postcode={75252 Paris}, 
            country={France}}

\begin{abstract}
High center-of-mass electromagnetic~(EM) interactions could produce decaying heavy leptons and hadrons, leading to neutrino generation. These processes might occur in the most extreme astrophysical scenarios, potentially altering the expected gamma-ray and neutrino fluxes in both the hadronic and the leptonic pictures. For instance, neutrinos could arise from high-redshift EM cascades, triggered by gamma rays beyond $10^{18} \; \text{eV}$ scattering background photons, from radio to ultraviolet energy bands. Such energetic gamma rays are predicted in cosmogenic models and in scenarios involving non-standard physics. On astrophysical scales, leptonic production of neutrinos could take place in active galactic nuclei cores, where several-TeV gamma rays interact with the X-ray photons from the hot corona. We explore these scenarios within the CRPropa Monte Carlo code framework, developing dedicated tools to account for leptonic production and decay of heavy leptons and hadrons. In particular, the latter are performed by interfacing with the PYTHIA event generator. With these novel tools, we characterise the spectrum and flavour composition of neutrinos emerging from cosmological EM cascades and from leptonic processes in the core of active galactic nuclei. Finally, we investigate the leptonic production of neutrinos in the context of the IceCube detection of NGC~1068.
\end{abstract}


\end{frontmatter}

\section{Introduction}

Propagating towards our observatories or in violent astrophysical environments, the most energetic charged particles -- electrons, protons and heavier nuclei, known as cosmic rays -- might interact with particle backgrounds, generating a cascade of secondary particles and, ultimately, gamma-ray and/or neutrino fluxes. Cascades are commonly classified as leptonic or electromagnetic~(EM) and hadronic, depending on interactions and particles at play. In the former, neutrino production is not expected, whereas the latter gives rise to interconnected  fluxes of gamma rays and neutrinos. Therefore, the detection (or absence) of neutrinos is an important benchmark for inferring the presence (or absence) of hadronic interactions, either during propagation or at the sources themselves. Investigating these scenarios could provide insights into the mechanisms capable of generating the most energetic gamma rays and neutrinos.   

Nevertheless, although extremely challenging in astrophysical contexts, neutrinos could still originate indirectly from leptonic interactions at high center-of-mass energies, without requiring the presence of any protons or nuclei. In fact, if kinematically allowed, heavy leptons or even hadrons may be produced through EM processes, with neutrinos subsequently emerging from their decays. For instance, extremely-high energetic gamma rays from cosmological sources could interact with background photons pervading intergalactic space, triggering EM cascades rich in muons and potentially tauons and/or hadrons as intermediate states. 

Among the most energetic gamma-ray and neutrino fluxes predicted within ``standard'' astrophysical scenarios are the so-called \textit{cosmogenic} fluxes, e.g.~\cite{greisen1966end, Zatsepin:1966jv, 1969PhLB...28..423B, stecker1973ultrahigh, alves2019open}. \textit{Cosmogenic} gamma rays could serve as potential projectiles for leptonic neutrino production during their cosmological propagation. However, the observed gamma-ray fluxes suffer from attenuation and/or ``deflection'' due to the generation of EM cascades initiated by interactions with cosmological background photons and the presence of magnetic fields (see e.g.~\cite{zdziarski1989absorption, Alves_Batista_2021}). 

Given the energy range of the densest background photon fields -- extending up to the ultraviolet band -- the production of muons via photon-photon interactions would require primary gamma rays with energies exceeding $\sim 10^{18} \; \text{eV}$, up to $\sim 10^{21}$--$10^{22} \; \text{eV}$ in the most optimistic cases~\cite{thompson2011upper}. Considering high-redshift sources, the gamma-ray energy threshold for muon pair production on the CMB moves down to lower values. Previous works have explored highly-energetic EM cascades developing over cosmological distances ($\gtrsim 100 \; \text{Mpc}$), determining the spectral characteristics of the outgoing neutrinos according to the source redshift and energy~\cite{li2007eevneutrinosassociateduhecr, Wang_2017, esmaeili2022ultrahigh, esmaeili2024neutrinos}. 

Gamma rays with energies of tens of~TeV may produce muons through interactions with $\sim \text{keV}$ target photons. This process may contribute to neutrino production in the hot corona of active galactic nuclei (AGNs)~\cite{hooper2023leptonic, latcollaboration2025fermidetectiongammarayemissionshot}, environments rich in X-ray radiation and potentially traversed by TeV~photons. This mechanism is one among the several proposed hadronic explanations~\cite{murase2020hidden, blanco2023neutrino, yasuda2024neutrinos, das2024revealing} for the detection of neutrinos from NGC~1068, a Compton-thick Seyfert~2 galaxy~\cite{urry1995unified, malizia2009fraction, pal2022x}, by the IceCube Neutrino Observatory~\cite{aartsen2020time, 2022}. Moreover, strong X-ray emissions from its core have been observed by NuSTAR~\cite{Marinucci_2015} and XMM-Newton~\cite{Bauer_2015}. The non-detection of TeV gamma rays from this source by MAGIC~\cite{Acciari_2019} may be a hint in favour of a thick source environment opaque to TeV~gamma rays, while Fermi-LAT detected gamma-ray fluxes up to $10 \; \text{GeV}$~\cite{Ajello_2020, Abdollahi_2020}. Despite the feasibility of the process within NGC~1068, a fully leptonic production of neutrinos could lead to tensions with X-ray and gamma-ray measurements of the source~\cite{das2024revealing}.    

Going beyond ``standard'' astrophysical frameworks, extremely-energetic fluxes of gamma rays and neutrinos are predicted by super heavy dark matter models~\cite{berezinsky1997ultrahigh, birkel1998extremely, kalashev2016constraining, he2020expected, esmaili2021first, das2023revisiting}, Z-burst scenarios~\cite{weiler1982resonant, fargion1999ultra}, topological defects~\cite{hindmarsh1995cosmic, Vachaspati_2015}, and from ``memory burden'' effects from primordial black holes~\cite{Thoss_2024, Chianese_2025, chianese2025highenergygammarayemissionmemoryburdened}. 

The predicted gamma-ray and neutrino fluxes, when combined with the observations from state-of-art high-energy experiments, could provide valuable constrains on both standard and exotic models. Systematic comparisons with current and upcoming neutrino observations, especially beyond several~PeV, are therefore crucial. New facilities have just started to observe the neutrino sky in the highest energy bands, marking significant steps toward unveiling the mechanisms underlying their production. Among the Cherenkov ground-based neutrino experiments, the IceCube Neutrino Observatory~\cite{williams2020status} and the KM3NeT~\cite{margiotta2023km3net} possess the highest detection capabilities. Through different techniques, the Pierre Auger Observatory~\cite{2015} can detect neutrinos fluxes in slightly higher energy ranges~\cite{Aab_2019}. Furthermore, a new class of radio telescopes could detect extremely-energetic neutrinos leveraging the Askaryan effect~\cite{dagkesamanskii1989radio}, such as the NuMoon experiment~\cite{Scholten_2006, scholten2009first}, a lunar-orbit radio observatory. 

In this paper, we present a novel approach to investigate scenarios of leptonic neutrino production, considering both source-intrinsic mechanisms and those occurring during propagation. To compute the neutrino fluxes from EM cascades, we develop dedicated plug-ins that complement the CRPropa code~\cite{Batista_2016, batista2022crpropa}. Notably, one of these extension enables the treatment of decays of secondary particles through an interface with the PYTHIA event generator~\cite{bierlich2022comprehensive}. This work may also serve as a proof of concept for exploring the occurrence of leptonic processes in various astrophysical and cosmological contexts, highlighting the importance of modelling the surrounding environments.

The properties of extremely-energetic EM cascades and the emerging neutrinos, together with their implementation in the CRPropa code, are described in Sec.~\ref{secExtrEMcascade}. In Sec.~\ref{secCosmologicalCascade}, after describing the relevant characteristics of extragalactic environments, in particular the background photons from the lowest radio to the ultraviolet energy bands, we investigate a scenario in which neutrinos are produced from cosmological monochromatic gamma-ray sources. Neutrino observables from simulations of these high-redshift sources, which emit gamma rays with energies extending from $10^{19}$ to several $10^{23} \; \text{eV}$, are presented in Sec.~\ref{secMonoSources}. The leptonic production of neutrinos in AGN cores is investigated in Sec.~\ref{SecNuFromAGN}, followed by a comparison with previous computations and neutrino detections from NGC~1068. A broader discussion of the results is provided in Sec.~\ref{secDisc}, leading to the conclusions in Sec.~\ref{secConcl}.   

\section{Neutrinos from leptonic processes}\label{secExtrEMcascade}

The production of heavy leptons or hadrons via interactions of gamma rays or energetic electrons and softer photons requires high center-of-mass energies. From their decays, neutrinos, along with other secondary energetic particles, arise. In the following, we outline the interaction and decay processes considered in this work, as well as their implementation in the development of the CRPropa plug-ins.

\subsection{Interactions}\label{interactionSubSec}
  
The interactions of a gamma ray with a background photon $\gamma_\text{bkg}$ could produce pairs of leptons, pions, kaons or higher mass hadrons~\cite{brodsky1981large, berger1987photon,nakazawa2008measurement,lu2020pi}, according to the kinematic threshold $s_\text{thr} \geq 4 m^{2}c^{4}$, with $m$ denoting the mass of the generated particles. The interactions that will be incorporated in our simulations are:  
\begin{itemize}[noitemsep,nolistsep]
    \item pair production~\cite{breit1934collision, nikishov1961absorption, gould1966opacity}: $\gamma+\gamma_\text{bkg} \to e^{-}+e^{+} $
    \item double pair production~\cite{cheng1970cross,brown1973role}: $\gamma+\gamma_\text{bkg} \to  2e^{-}+2e^{+}$ 
    \item muon pair production~\cite{akhiezer1953quantum, Athar_2001}: $\gamma+\gamma_\text{bkg} \to \mu^{-}+\mu^{+} $
    \item charged pion pair production~\cite{akhiezer1953quantum,lyth1971theoretical,morgan1987low,tasso1986vector,boyer1990two,nakazawa2005measurement}: $\gamma+\gamma_\text{bkg} \to \pi^{-}+\pi^{+} $
    \item tauon pair production~\cite{akhiezer1953quantum}: $\gamma+\gamma_\text{bkg} \to \tau^{-}+\tau^{+}$
\end{itemize}
We develop original plug-ins suitable for introducing in the CRPropa framework the production of heavy leptons and pions from photon-photon interactions. They are \href{https://github.com/GDMarco/EMMuonPairProduction}{\texttt{EMMuonPairProduction}}, \href{https://github.com/GDMarco/EMTauonPairProduction}{\texttt{EMTauonPairProduction}} and \href{https://github.com/GDMarco/EMChargedPionPairProduction}{\texttt{EMChargedPionPairProduction}}. The structure of these plug-ins resembles the usual style of CRPropa interaction modules.

If kinematically allowed, hadrons with higher masses than the ones quoted above could be produced by photon-photon interactions, but we have not included these in our analysis. Thus, final states as $K^{-}K^{+}$, $\pi^{0}\pi^{0}$, $q\bar{q}$, $\pi^{0}\eta$, and so on~\cite{nikishov1961absorption, gould1966opacity, lu2020pi} could arise from photon-photon interactions. These occur at much higher center-of-mass energies compared to muon or pion pair productions, thus it does not significantly affect our results. Furthermore, electron-positron pairs could be produced by gamma rays moving through magnetised regions, via $\gamma+B_\text{ext} \to e^{-}+e^{+}$, where $B_\text{ext}$ is the external magnetic field. 

Energetic electrons, e.g. either accelerated in extreme sources or produced in EM cascades, may scatter background photons producing either gamma rays or lepton/charged hadron pairs. In this work, we include: 
\begin{itemize}[noitemsep,nolistsep]
    \item inverse Compton scattering \cite{jones1968calculated}: $e+\gamma_\text{bkg} \to e+\gamma$ 
    \item triplet pair production \cite{motz1969pair}: $e+\gamma_\text{bkg} \to e+e^{-}+e^{+}$
    \item electron-muon pair production \cite{abarbanel1968low}: $e+\gamma_\text{bkg} \to e+\mu^{-}+\mu^{+}$
\end{itemize}
The latter process is implemented in our simulation code through the \href{https://github.com/GDMarco/EMElectronMuonPairProduction}{\texttt{EMElectronMuonPairProduction}} plug-in. Furthermore, these energetic electrons, as well as other charged particles, undergo deflections and synchrotron losses, when going through magnetised astrophysical environments. 

The cross sections of the processes described above are reported in~\ref{appEMprocCS}, together with more details on the plug-ins for the CRPropa code. 

\subsection{Decay of leptons and mesons}

If a leptonic high center-of-mass interaction occurs, it could result in the production of \textit{unstable} heavy leptons or mesons, which propagate over short distances before decaying. The decay processes useful for our purposes are:  
\begin{itemize}[noitemsep,nolistsep]
    \item $\mu^{-} \to e^{-}+\bar{\nu}_{e}+\nu_{\mu}$
    \item $\pi^{-} \to \mu^{-}+\bar{\nu}_{\mu}$
    \item $\tau^{-} \to \nu_{\tau} + \text{leptons / mesons} + ...$ 
\end{itemize}
Thus, neutrinos with different flavours and energies, with additional secondary particles, arise from the decays of muons, tauons and charged pions. These decays are performed in CRPropa through the development of the dedicated \href{https://github.com/GDMarco/CRPYTHIAxDecays}{\texttt{CRPYTHIAxDecays}} plug-in. It exploits some functionalities of the PYTHIA code~\cite{bierlich2022comprehensive}. The general idea of the plug-in is to perform decays according to particle lifetimes and, then, generate the products through the PYTHIA code. These products are re-injected in the CRPropa simulation. Users can also choose if to apply decay angular corrections, mostly negligible for very energetic particles. Physical and technical details on the decay plug-in are treated in~\ref{appDecSpec}. 

\section{Neutrinos from cosmological propagation of highly-energetic gamma rays}\label{secCosmologicalCascade}

In this section, descriptions of the extragalactic environment and of the theory of gamma-ray propagation are provided, respectively in Secs.~\ref{secExtraEnv} and~\ref{subsecCosmInt}. The simulations of propagating highly-energetic gamma rays from generic monochromatic sources at high redshifts are constructed based on current knowledge of the extragalactic environments and the interaction processes described in Sec.~\ref{secExtrEMcascade}. In Sec.~\ref{secMonoSources}, we characterise the observed neutrinos emerging from the interaction of such energetic gamma rays with low-energy photon backgrounds. 


\subsection{Extragalactic environment}\label{secExtraEnv}

The space between galaxies, clusters and filaments, to which we refer to as extragalactic space, is known to be imbued by magnetic and radiation fields, as well as neutrinos, gases and electrons. In this work, we focus on photon backgrounds, which serve as potential targets for interactions with high-energy gamma rays and electrons~\cite{gould1978subject, svensson1990photon}. The effects of extragalactic magnetic fields on the cascade development~\cite{Alves_Batista_2021} are neglected in this work. It would cause the deflection of the charged particles in the cascade, roughly evaluated through the usual Larmor radius formula, relating the particle energy ($E$) to the magnetic field intensity perpendicular to the particle direction of motion ($B_{\perp}$) as $R_{L}=E\cdot(ecB_{\perp})^{-1}$, wherein $e$ is the elementary charge and $c$ the speed of light. A brief excursus on extremely-energetic neutrino absorption on the cosmic neutrino background (C$\nu$B) is in~\ref{app_nuAbs}. Possible effects due to pervasive electrons and nuclei in the extragalactic space, e.g. plasma instabilities~\cite{alves2019impact, castro2024influence}, are ignored.   

\paragraph*{Photon backgrounds}
Low-energy background photons in extragalactic space range from the radio to ultraviolet frequencies. These photons are either relics from the early universe or by-products of the formation and evolution of structures, including stars, galaxies and clusters. The primary example of the former class is the cosmic microwave background (CMB),  composed of relic photons from the \textit{recombination} era, described by a black-body field. The CMB is currently peaked at a temperature of around $2.73 \; \text{K}$~\cite{aghanim2020planck}. 
The diffuse photons produced by astrophysical sources tend to compose the cosmic radio background (CRB)~\cite{protheroe1996new, fixsen2011arcade, nictu2021updated} and extragalactic background light (EBL)~\cite{gilmore2012semi, saldana2021observational, finke2022modeling}, the latter extending from the infrared through the visible spectrum.


\subsection{Cosmological interaction rates}\label{subsecCosmInt}
\begin{figure*}[ht]
    \centering
    \includegraphics[width=\textwidth]{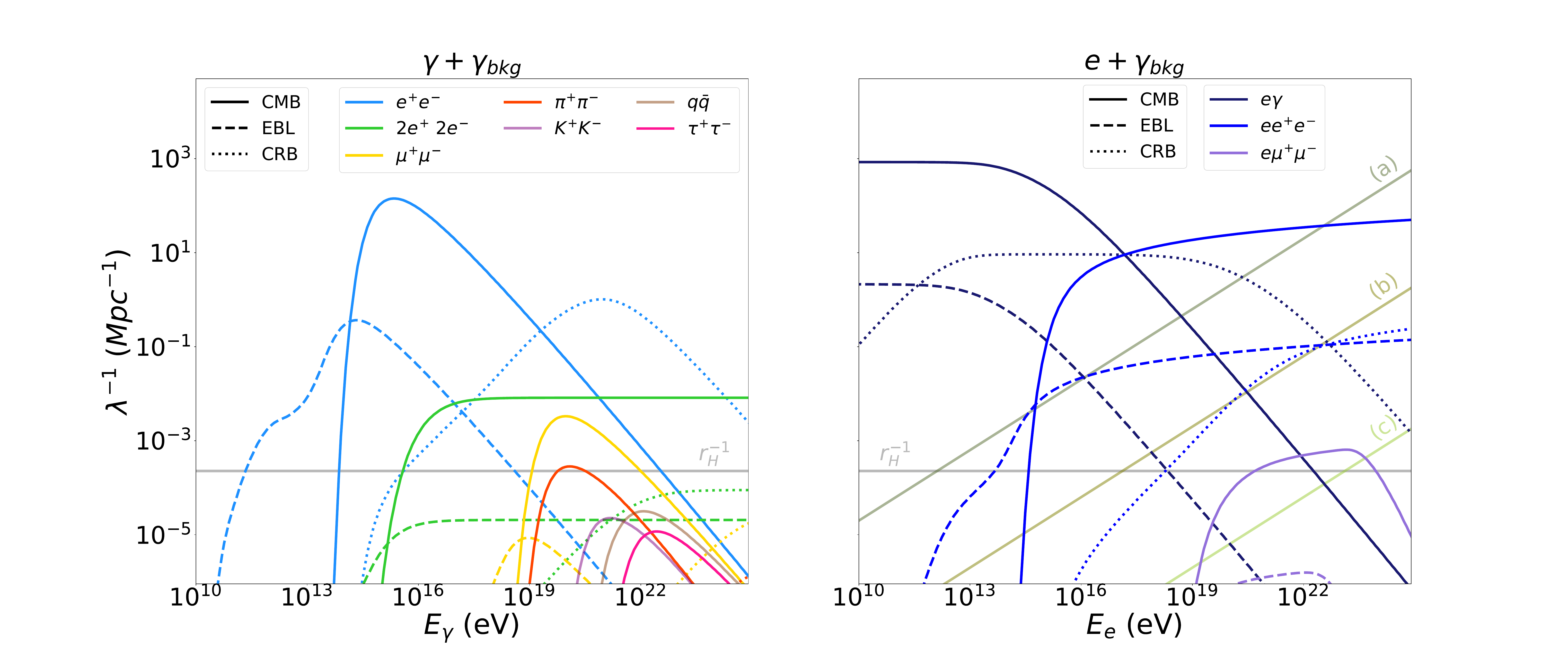}
    \caption{Inverse mean free paths as a function of the energy of the propagating gamma ray or electron. They are computed for the background densities at redshift z=0. The \textit{linestyle legend} refers to the background photons. The EBL model employed in the rate computations (\textit{dashed lines}) is the one of Ref.~\cite{saldana2021observational}, while the CRB one (\textit{dotted lines}) comes from Ref.~\cite{nictu2021updated}. The \textit{colour legend} indicates the interaction at play. The \textit{horizontal grey lines} stand for the inverse Hubble radius, i.e. $r_{H}^{-1}$. The \textit{left panel} refers to the photon-photon interaction rates, while the one on the \textit{right} for the electron-photon ones. The \textit{greenish diagonal lines} in the latter are the inverse average distances travelled by electrons propagating in randomly-oriented magnetic fields of intensity $B$ and domain size $l_{B}$. The line labelled as \textit{(a)} refers to $B = 10^{-10} \; \text{G}$ and $l_{B} = 10^{-1} \; \text{Mpc}$, \textit{(b)} to $B = 10^{-12} \; \text{G}$ and $l_{B} = 1 \; \text{Mpc}$, \textit{(c)} to $B = 10^{-14} \; \text{G}$ and $l_{B} = 10^{2} \; \text{Mpc}$. }
    \label{fig:intRatesCosmo}
\end{figure*}

Gamma rays produced by cosmologically distant sources traverse the extragalactic space pervaded by the soft background photons described in Sec.~\ref{secExtraEnv}, before reaching our observatories. Figure~\ref{fig:intRatesCosmo} shows the inverse mean free paths (i.e., interaction rates) for standard EM processes, namely pair production and inverse Compton, on cosmological background photons. Also shown are higher-order EM processes, namely double and triplet pair production, along with interactions leading to the production of heavy leptons or hadrons. The rates are juxtaposed with the inverse Hubble radius, computed for $H_{0} = 67.77 \; \text{km}\; \text{Mpc}^{-1}\;\text{s}^{-1}$~\cite{2014}. 

In EM cascades developing in the extragalactic environment, in which the magnetic field intensity is $B_{\text{ext}} \lesssim 10^{-10} \; \text{G}$~\cite{jedamzik2019stringent}, the synchrotron energy loss is negligible for electrons. It is shown in the inverse average synchrotron distances of electrons computed for various combinations of the extragalactic magnetic field parameter, the \textit{greenish lines} in the \textit{right} panel of Fig.~\ref{fig:intRatesCosmo}. Therefore, lifetimes of the \textit{unstable} leptons and hadrons are shorter than synchrotron loss timescales in such low-intensity magnetic fields. 

\subsection{Simulations}\label{secMonoSources}
\begin{figure}[ht!]
    \centering
    \includegraphics[width=0.49\textwidth]{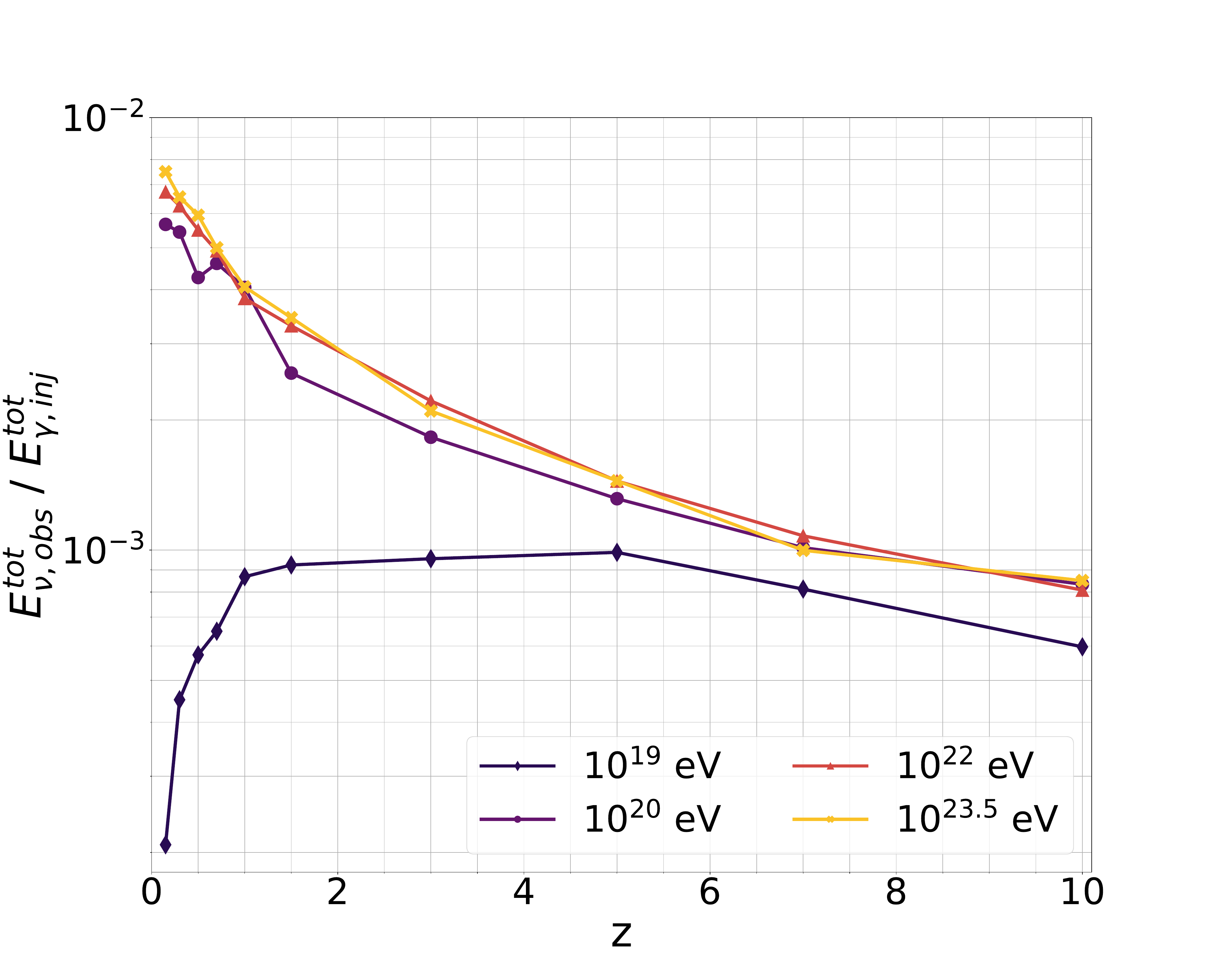}
    \caption{Ratio between the total observed neutrino energy and the total injected gamma-ray energy. The sources are characterised by the monochromatic energy indicated in the \textit{legend}, and the redshift, on the x-axis.}
    \label{fig:enFrac_inj}
\end{figure}

We simulate monochromatic gamma-ray sources with $E_{\gamma, \text{inj}} \in [10^{19}, \;10^{23.5}] \; \text{eV}$ to investigate the production of neutrinos during the development of EM cascades in the extragalactic environment. In this energy range, the interaction rate peaks for heavy lepton and meson production on the CMB are covered (Fig.~\ref{fig:intRatesCosmo}). The sources are placed at cosmological distances, conventionally corresponding to redshifts $\gtrsim 0.1$; thus, the redshifts of the investigated sources span from 0.15 to 10. The background photons incorporated in the simulations are the CRB model from Ref.~\cite{nictu2021updated}, the CMB, and the EBL from Ref.~\cite{saldana2021observational}. Figure~\ref{fig:enFrac_inj} shows the fraction of the injected gamma-ray energy ultimately transferred to neutrinos, with a detection threshold of $10^{16} \; \text{eV}$. The rest of the energy ends up in lower energy, down to the~GeV band, or ultimately absorbed gamma rays and electrons. The characterisation of the electromagnetic counterparts is addressed to future dedicated studies within this framework. For redshifts lower than 1, the lowest-energy sources convert less than the $0.1 \%$ of their initial energy into neutrinos. For higher redshifts, the neutrino energy fractions are close to $0.1 \%$, except for a slight decrease observed for the farthest source. For sources with energies up to $10^{23} \; \text{eV}$ the  descending profiles of the neutrino energy fraction with respect to the total gamma-ray energy are rather similar. The highest fractions are observed for the closest sources, viz. at $z=0.15$. 

For most of the sources studied, neutrinos produced via muons decays account for approximately $\sim 90-95\%$ of the total energy transferred into neutrinos. Interestingly, for the two highest-energy sources, this fraction falls below $\lesssim 90\%$, indicating a relatively higher contribution to the detected energy budget from neutrinos produced via charged pion and tauon decays. The EM cascades from the lowest-energy source start producing pions at redshifts $z \gtrsim 1.5$. Moreover, it is worth noting that, even at z=0, the gamma rays from the two highest-energy monochromatic sources might produce tauons by interacting with the CMB photons, despite the very low rates~(Fig.~\ref{fig:intRatesCosmo}). Tauons start to be systematically produced in EM cascades from distant sources with injected gamma-ray energies larger than $10^{21} \; \text{eV}$. Sources emitting gamma rays with energies $\gtrsim 10^{23} \; \text{eV}$ are the most efficient tauon factories across all redshifts. In these cases, the energy carried by tauon-produced neutrinos is always $\gtrsim 1 \%$ of the total observed neutrino energy.

\begin{figure}[htb!]
    \centering
    \includegraphics[width=0.49\textwidth]{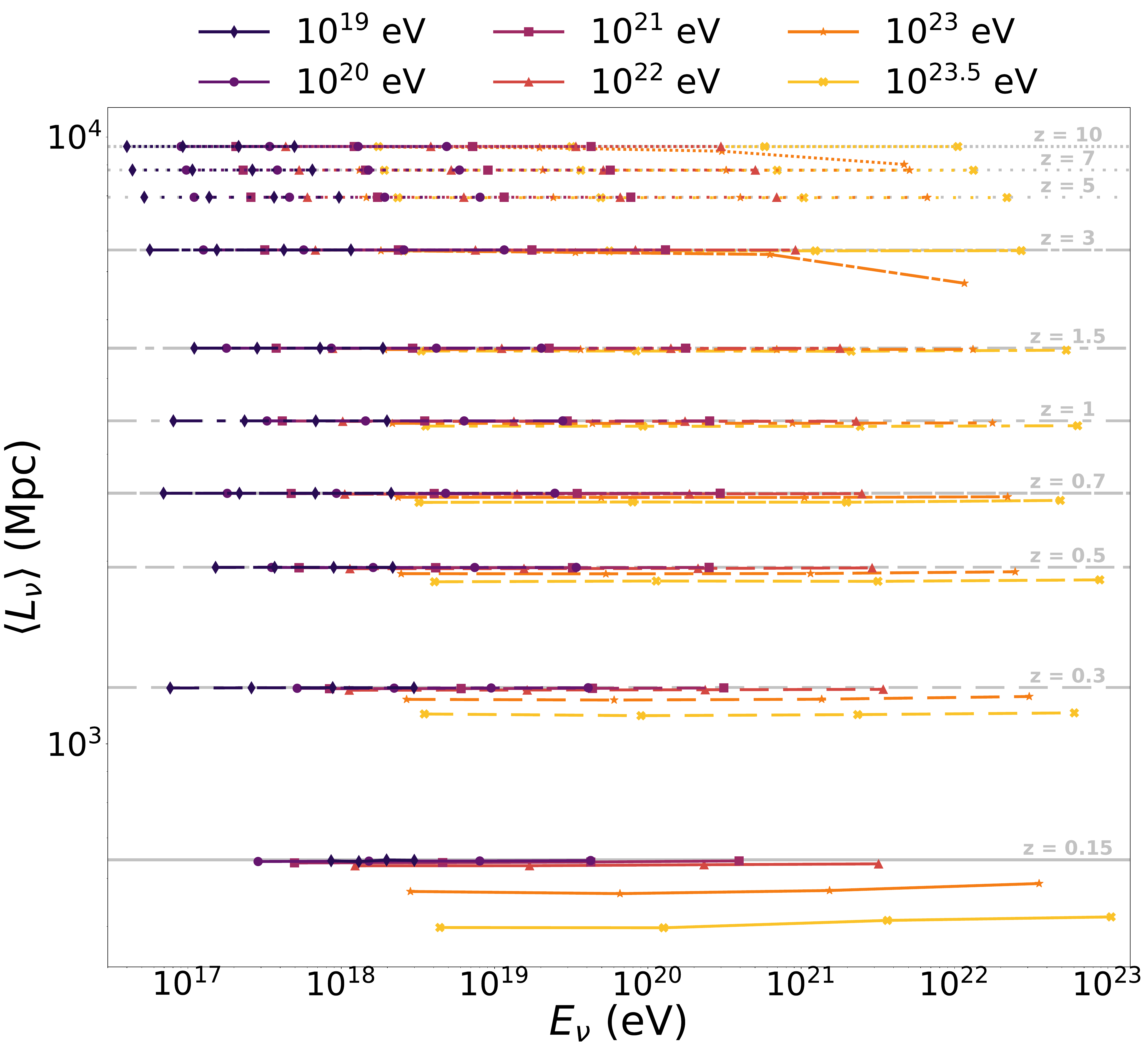}
    \caption{The average distance travelled by the observed neutrinos in dependence to the neutrino energies. In the \textit{upper legend}, the energies of the monochromatic gamma-ray sources are showed. The \textit{labelled grey lines} indicate the comoving distance corresponding to the source redshift. The various \textit{line styles} refer to the different redshifts of the sources.}
    \label{fig:LSdist}
\end{figure}

The distances the neutrinos travel before reaching Earth are shown in Fig.~\ref{fig:LSdist}, as grouped in four energy bins. Except for the two most energetic sources, the distances travelled by the neutrinos do not show any considerable dependence on the energy, since they are produced in the vicinities of the sources. In contrast, the energy--distance profiles for monochromatic sources with $E_{\gamma, \text{inj}} \gtrsim 10^{23} \; \text{eV}$ demonstrate that the neutrinos from low-redshift sources tend to be produced closer to the observer.   

\begin{figure*}[ht!]
    \centering
    \includegraphics[width=\textwidth]{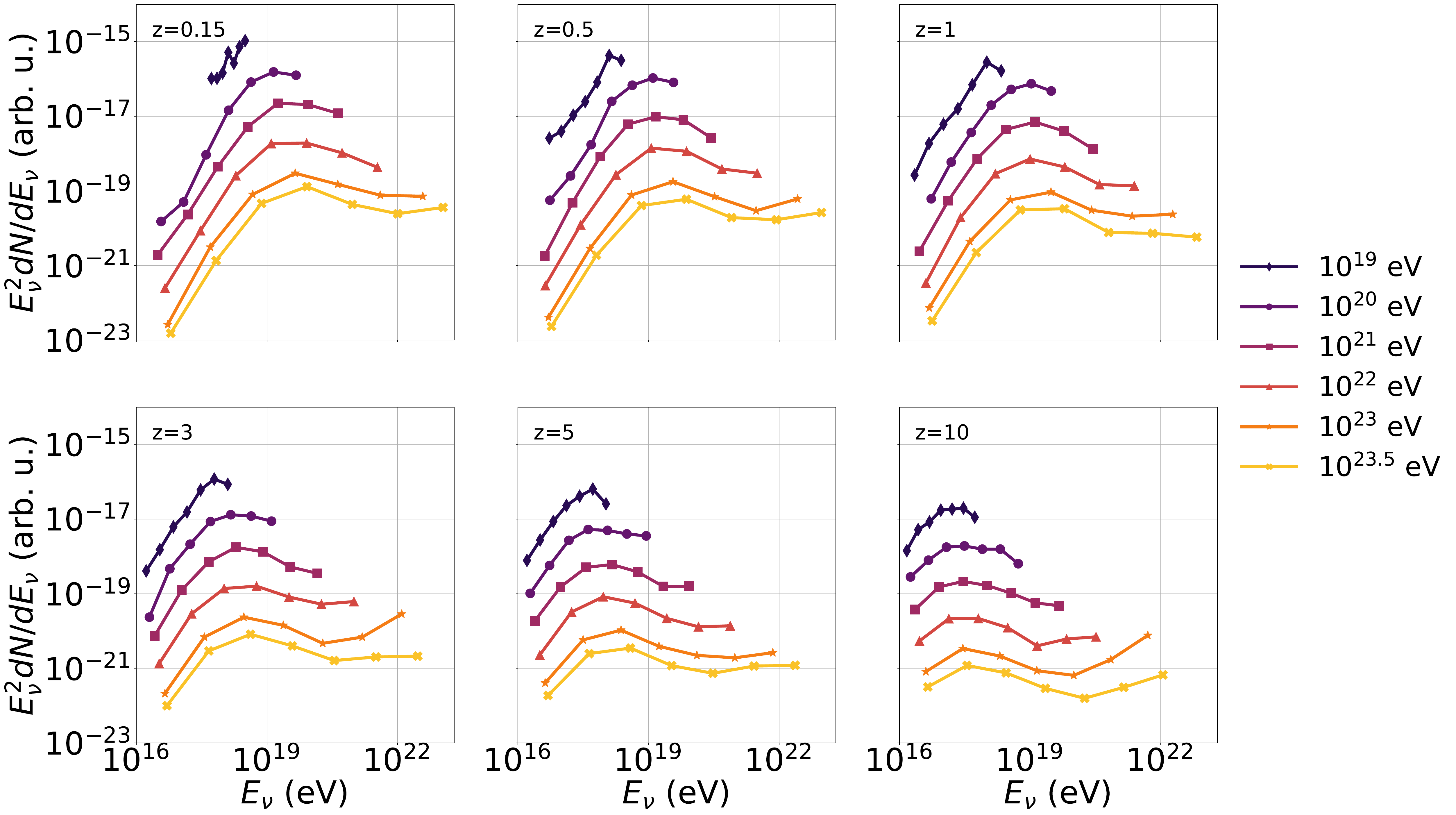}
    \caption{All-flavour neutrino energy spectra from the monochromatic gamma-ray sources at different redshifts. They are normalised to the total injected energies.} 
    \label{fig:nuSED}
\end{figure*}

\paragraph*{Neutrino fluxes} The all-flavour neutrino spectra arising from the propagation of the gamma rays injected by the various monochromatic sources are in Fig.~\ref{fig:nuSED}. At lower injection energies, the spectral profiles are broadly similar. The steep rises in the spectra span a broad energy range, from the simulation threshold up to approximately $10^{18}$–$10^{20} \; \text{eV}$, depending on both the source redshift and the injected energy. For a fixed injected energy, increasing the redshift shifts the spectral peaks to lower neutrino energies. Conversely, compared to sources at a fixed redshift, i.e., within a single quadrant of Fig.~\ref{fig:nuSED}, higher injected energies result in broader spectral peaks. Once the maximum is reached, a moderate decline in the profiles follows for sources with $E_{\gamma, \text{inj}} \gtrsim 10^{20} \; \text{eV}$, then either smoothly rising again in the case of the most energetic sources, or remaining flat in the upper energy bins. 

The most energetic neutrinos observed from each source depends mainly on the source distance. In general, the distance determines the redshift dependence of the observed maximum energy, following a factor $(1+z)^{-1}$, due to the adiabatic energy losses caused by the universe expansion. In our scenario, this factor is at most $\sim 10$, considering the farthest simulated sources, thus reducing the observed energy by, at most, one order of magnitude. In fact, the energy ratios between the most energetic observed neutrino and the primary gamma ray range from $60$ to $70\%$, except for $10^{19} \; \text{eV}$ sources in which it is below $45\%$, for the closest sources, down to $10\%$ for the farthest ones. The redshift dependence of the ratios follows a hyperbolic trend, reflecting the adiabatic energy loss, in line with the profiles of Fig.~\ref{fig:enFrac_inj}. 

\paragraph*{Flavour composition} Without flavour oscillations, the percentages of electronic neutrinos lie in the range $45-51\%$, while the muonic neutrinos account for approximately $49-54\%$. The fractions of tauonic neutrinos are appreciable only for the most energetic sources studied. They are found to be lower than approximately $0.2\%$.

As shown in Fig.~\ref{fig:LSdist}, the neutrinos produced in the EM cascades travel distances greater than several tens of Mpc. Assuming decohered flavour states, $\theta_{\odot} \sim 34^{\circ}$ solar mixing angle\footnote{Defined in the context of the standard three-flavour picture, which assumes small ``solar" masses splitting and large ``atmospheric" splitting~\cite{esteban2025nufit}.}~\cite{esteban2025nufit} and large mixing phases, the energy and distance independent oscillation probabilities computed in Ref.~\cite{anchordoqui2014cosmic} could be applied to the prompt compositions, resulting in $(\nu_{e} : \nu_{\mu} : \nu_{\tau}) \sim (0.38:0.31:0.31)$.

\begin{figure*}[t]
  \centering

  \begin{subfigure}[b]{0.48\textwidth}
    \centering
    \includegraphics[width=\linewidth]{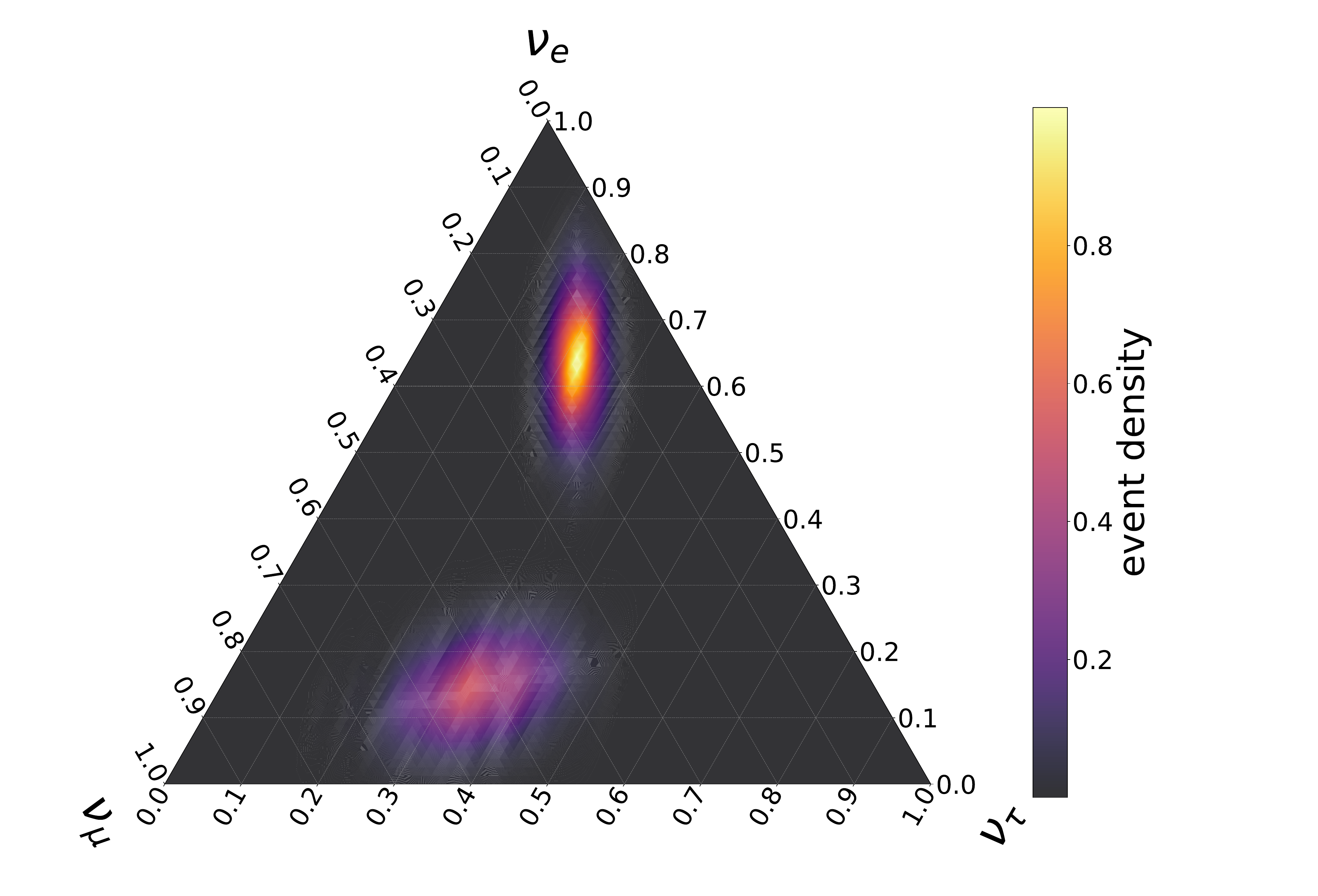}
    \caption{$E_{\gamma,\text{inj}}=10^{20}\,$ eV, $z=10$}
    \label{fig:oscTr1e20_10}
  \end{subfigure}\hfill
  \begin{subfigure}[b]{0.48\textwidth}
    \centering
    \includegraphics[width=\linewidth]{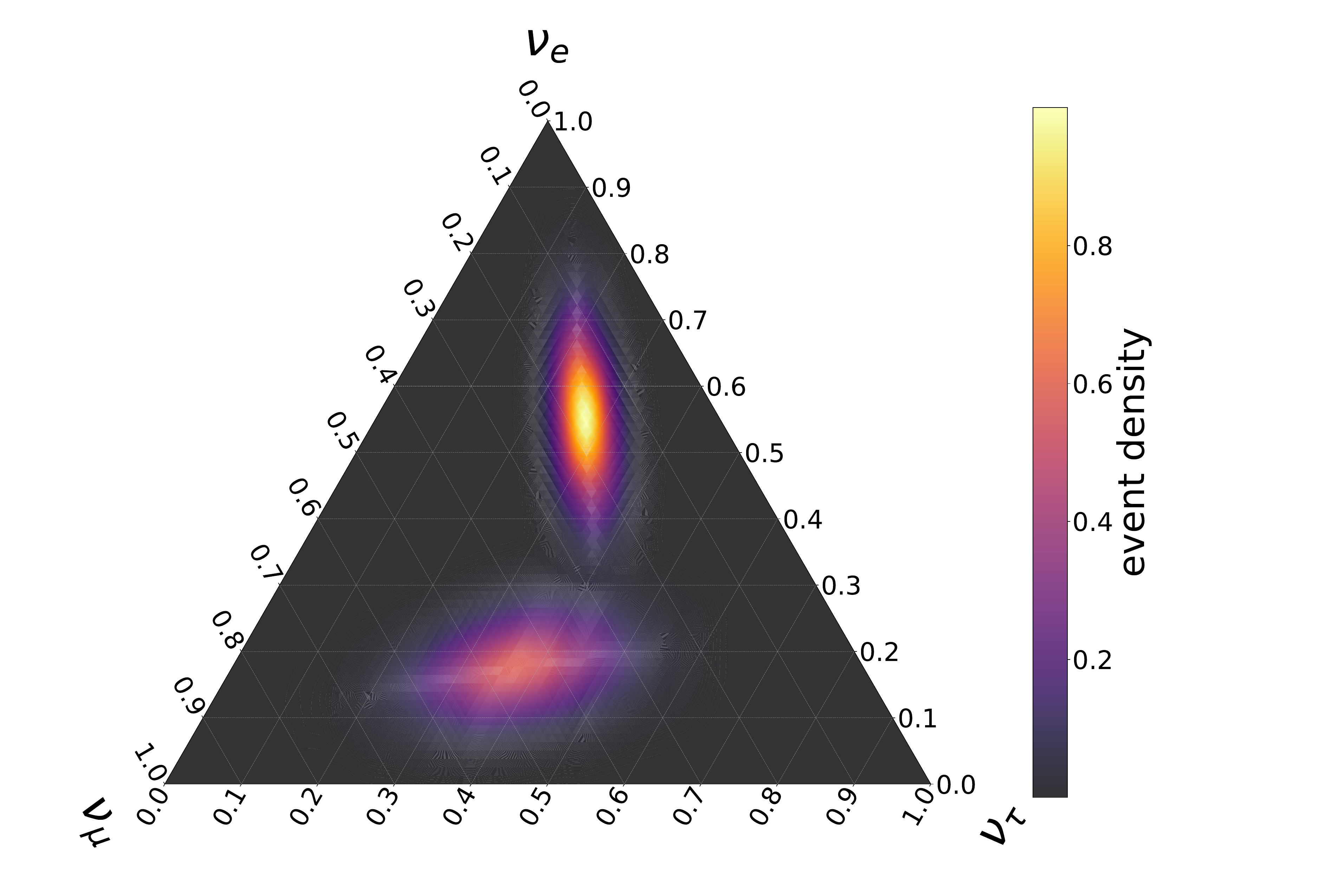}
    \caption{$E_{\gamma,\text{inj}}=10^{22}\,$eV, $z=7$}
    \label{fig:oscTr1e22_7}
  \end{subfigure}

  \vspace{3mm}

  \begin{subfigure}[b]{0.48\textwidth}
    \centering
    \includegraphics[width=\linewidth]{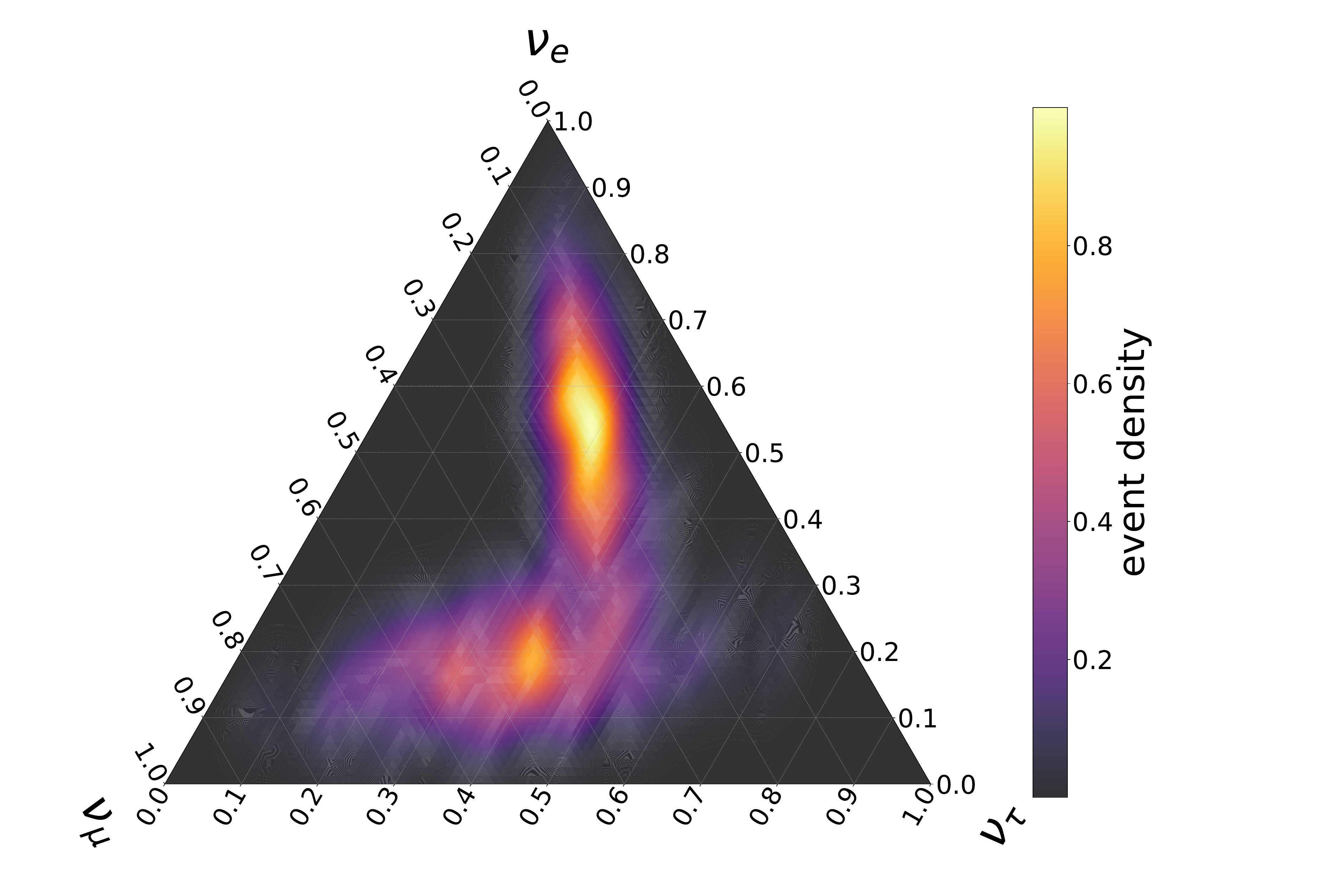}
    \caption{$E_{\gamma,\text{inj}}=10^{20}\,$eV, $z=0.15$}
    \label{fig:oscTr1e20_015}
  \end{subfigure}\hfill
  \begin{subfigure}[b]{0.48\textwidth}
    \centering
    \includegraphics[width=\linewidth]{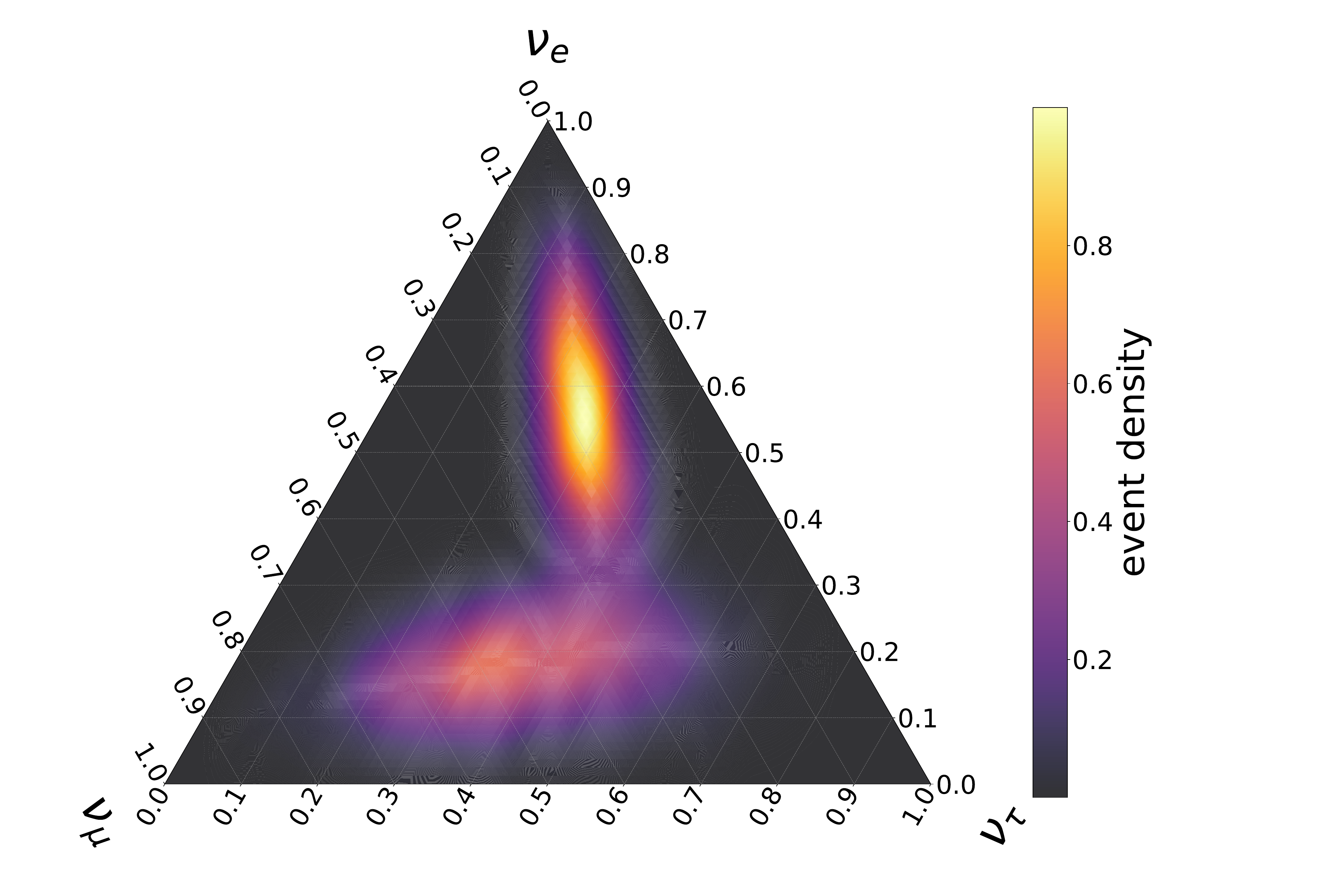}
    \caption{$E_{\gamma,\text{inj}}=10^{22}\,$eV, $z=0.15$}
    \label{fig:oscTr1e22_015}
  \end{subfigure}

  \caption{Triangular flavour plot after applying propagation of exact oscillation probabilities.}
  \label{fig:cosmoTriFlav}
\end{figure*}

By computing the exact oscillation probabilities, which depends on the neutrino energy and the distance travelled, as explained in~\ref{app_propProb}, the expected flavour distributions of the observed neutrinos are shown in the triangle plots of Fig.~\ref{fig:cosmoTriFlav}. These correspond to the most distant source ($z=10$) with an injection energy of $10^{20} \; \text{eV}$~(Fig.~\ref{fig:oscTr1e20_10}) and to the source of $10^{22} \; \text{eV}$ placed at $z=7$~(Fig.~\ref{fig:oscTr1e22_7}). Both plots exhibit two \textit{dense} probability regions: one more likely electronic and the other favouring muonic neutrinos. In both regions, the tauonic flavour probabilities lie between 0.3 and 0.6. The corresponding flavour plots for closest sources are presented in Figs.~\ref{fig:oscTr1e20_015} and~\ref{fig:oscTr1e22_015}. They exhibit broader and less well-defined flavour density regions due to the reduced distance travelled.   

\section{Neutrinos from leptonic processes in AGN coronae}\label{SecNuFromAGN}
The hot coronae of AGNs have been proposed as suitable sites in which neutrinos could be produced through the EM processes treated in Sec.~\ref{secExtrEMcascade}~\cite{hooper2023leptonic}. This model has been specifically applied to the case of the neutrinos detected from NGC~1068~\cite{2022}. In X-ray--luminous AGNs, usually classified as Seyfert galaxies, $\text{keV}$ photons originating in the corona~\cite{sunyaev1980comptonization, murase2020hidden, latcollaboration2025fermidetectiongammarayemissionshot} could act as targets for gamma rays produced either by synchrotron emission from electrons in the torus~\cite{lopez2020alma} or from shocks within the corona itself~\cite{Inoue_2019, Inoue_2020}. 

\begin{figure}[htb]
    \centering
    \includegraphics[width=0.49\textwidth]{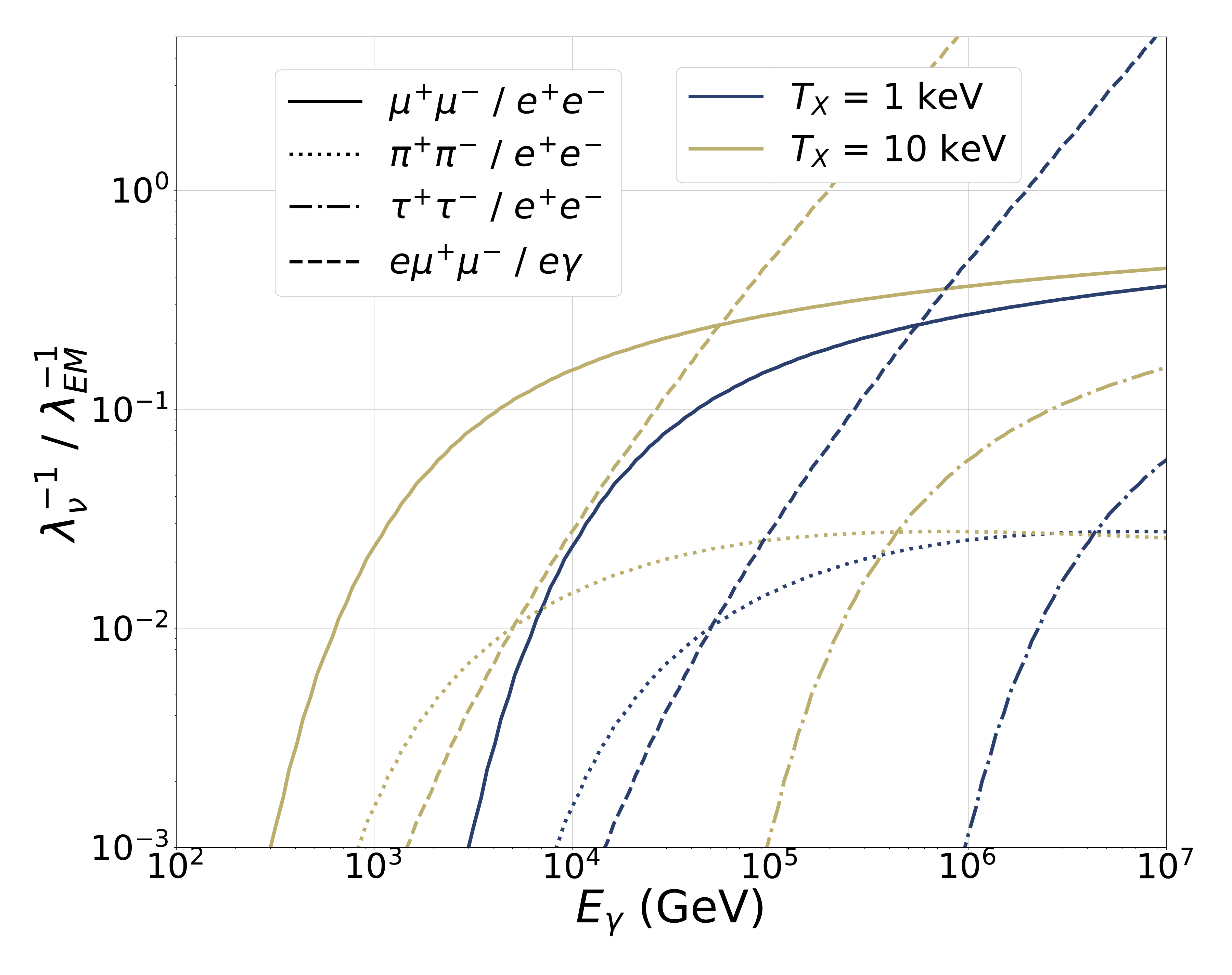}
    \caption{Interaction rate ratios in which $\lambda^{-1}_{\nu}$ refers to the processes that produces heavy leptons or mesons, thus neutrinos at the end, while $\lambda^{-1}_{\text{EM}}$ to the ones that result in either gamma-ray or electron productions. The \textit{left-side linestyle legend} indicates which processes are involved, specifying the final states. The \textit{central color legend} indicates the characteristic temperature of the X-ray field.}
    \label{fig:intRatio_XrCorona}
\end{figure}

Pair production on X-ray photons of energy $\epsilon_{X}$ is kinematically allowed if the energy of the gamma ray is
\begin{equation}
    E_{\gamma} \geq \dfrac{4 \, m^{2}c^{4}}{\epsilon_{X}(1-\cos{\theta})} \,,
\end{equation}
where $\theta$ is the coplanar scattering angle between the two momenta. Thus, for head-on collisions, considering $\epsilon_{X} = 10 \; \text{keV}$, muon pairs could be produced by gamma rays of energy $E_{\gamma} \gtrsim 300 \; \text{GeV}$. The interaction thresholds are reflected in the rate ratios, viz. the \textit{solid}, \textit{dotted}, and \textit{dash-dotted lines} of Fig.~\ref{fig:intRatio_XrCorona}, assuming that the X-ray photons are distributed as a blackbody with temperature $T_{X}$. The assumption on the corona X-ray spectral distribution could be refined by adopting a power-law spectral model with a possible cut-off, as suggested by observations~\cite{Trakhtenbrot_2017, Ricci_2018}. To produce such energetic gamma rays, regions of moderately large magnetic fields, i.e. $\mathcal{O}(\text{kG})$, are required~\cite{hooper2023leptonic}. In an alternative picture, gamma-ray production may also arise from accelerated electrons scattering off soft photons from the accretion disc~\cite{koratkar1999ultraviolet, liu2017centrally}. In this framework, strong magnetic fields are not required, and inverse Compton scattering (see Sec.~\ref{interactionSubSec}) by hot corona electrons could be an efficient producer of multi-TeV gamma rays. 

Assuming an X-ray dense corona that is optically thick to gamma rays, an EM cascade can be triggered and potentially develop. As an order-of-magnitude estimate, the radius of NGC~1068 corona is approximately $10^{-3} \; \text{pc}$~\cite{murase2020hidden}. As shown in Fig.~\ref{fig:intRatio_XrCorona}, the ratio between muon and electron pair productions become larger than $0.1$ for $E_{\gamma} \gtrsim 30 \; (3) \; \text{TeV}$, assuming a characteristic corona temperature of $T_{X} = 1 \; (10) \; \text{keV}$. In contrast, the production of charged pion or tauon pairs remains subdominant due to their higher kinematic thresholds. Electrons produced in the cascade could, in turn, interact with X-ray photons producing either gamma rays or muon pairs -- the ratios are the \textit{dashed lines} of Fig.~\ref{fig:intRatio_XrCorona}.

For $T_{X} = 10 \; \text{keV}$, the muon pair production rate from hundreds-of-TeV electrons starts to dominate over inverse Compton scattering one. Nevertheless, such energetic electrons suffer from severe radiative cooling owing to the high magnetic fields ($\gtrsim \text{kG}$) of the plasma in the corona~\cite{jiang2014radiation, murase2020hidden, Murase_2023, fiorillo2025neutrinostxs0506056coronal} and within the possible clumps~\cite{Khangulyan_2021}. Moreover, the high magnetic-field intensities could prevent the produced muons and charged pions from decaying before losing a substantial amount of their energy through synchrotron emission. 

These considerations make the treatment of the cascade electrons propagating within the corona a delicate point. Moreover, the presence of softer photons in the corona, e.g., from the accretion disc~\cite{koratkar1999ultraviolet, murase2020hidden}, could lead to further gamma-ray absorption solely through electron pair production. This effect depends on the photon densities and their spatial distribution within the corona itself. 

\begin{figure}[htb!]
    \centering
    \includegraphics[width=0.49\textwidth]{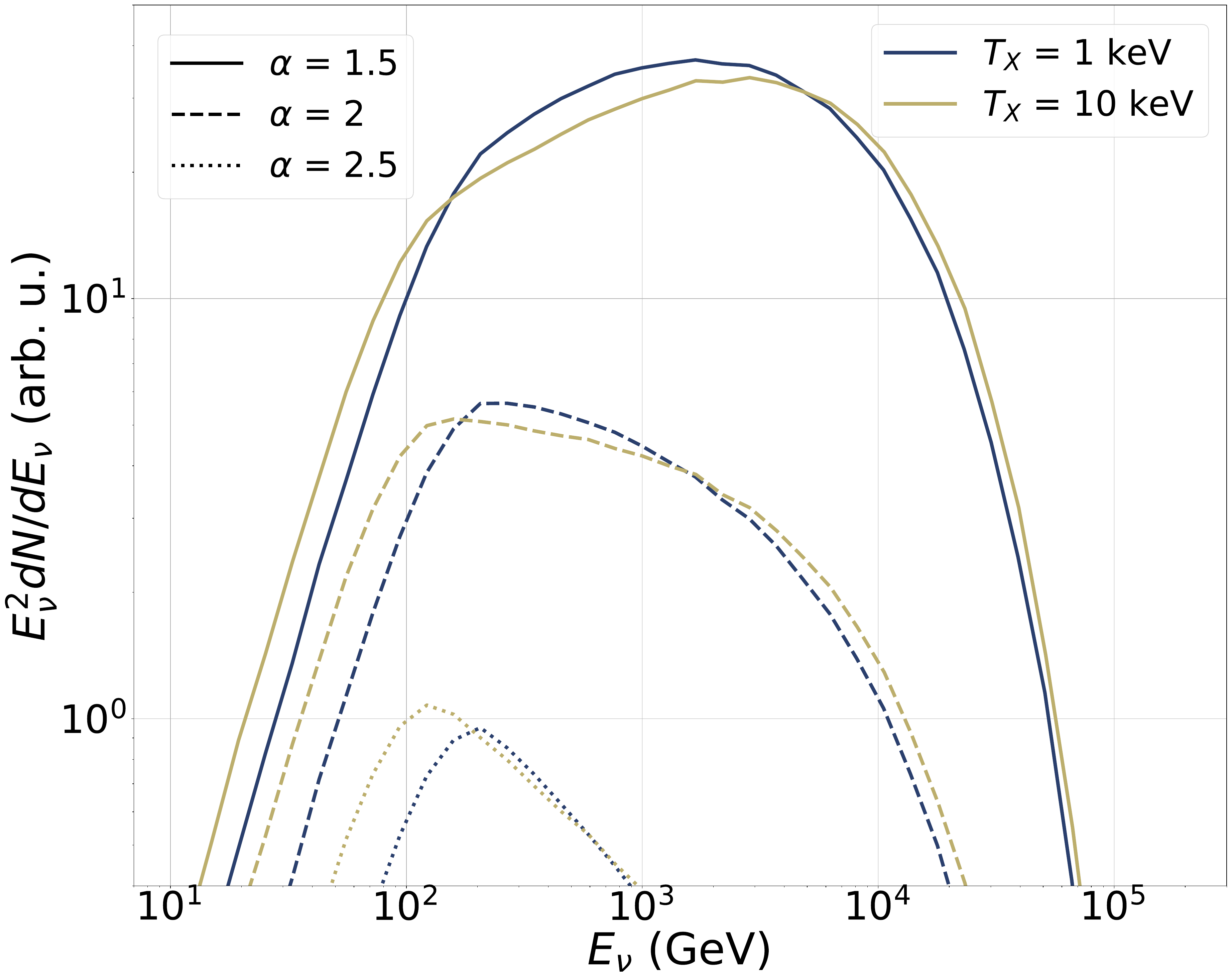}
    \caption{All-flavour neutrino spectra, assuming different gamma-ray spectral indexes and X-ray background characteristic temperatures. The fluxes are normalised to the prompt gamma-ray spectra. The energy cut-off $E_{\text{max}}$ of Eq.~\ref{power-lawCOEq} is set to $20 \; \text{TeV}$.}
    \label{fig:spectraAGNnu}
\end{figure}

The prompt gamma-ray emission of this work is modelled as an exponential cut-off power-law phenomenological profile, resembling synchrotron production of gamma rays from the corona electrons~\cite{esmaeili2024neutrinos}. It follows the expression: 
\begin{equation}\label{power-lawCOEq}
\frac{\mathrm{d}N}{\mathrm{d}E}\propto E^{-\alpha} \exp\left(-\dfrac{E}{E_{\text{max}}}\right) \,,
\end{equation} 
where $\alpha$ is the spectral index and $E_{\text{max}}$ the cut-off energy. Including only electron and muon pair production processes, the resulting neutrino energy spectra resulting for different choices of $\alpha$ and $T_{X}$ are depicted in Fig.~\ref{fig:spectraAGNnu}. As in Refs.~\cite{hooper2023leptonic, esmaeili2024neutrinos}, the cutoff on the prompt gamma-ray spectra is set to $20 \; \text{TeV}$. This cutoff approximates the scenario in which gamma rays are produced via electron synchrotron emission in a magnetic field of $5 \; \text{kG}$. In this computation, the source redshift is set to $z\sim 0.004$\footnote{It is the fiducial redshift of NGC~1068, as reported in the \href{https://ned.ipac.caltech.edu/}{NASA/IPAC Extragalactic Database}.}.  

The gamma-ray spectral index of Eq.~\ref{power-lawCOEq} is related to the electron power-law distribution, governed by the index $p$, through the relation $\alpha = \frac{p+1}{2}$, where typically $p \sim 2$ in AGN studies~\cite{raginski2016agn, hooper2023leptonic}. Harder spectra, i.e. lower spectral indexes, produce higher neutrinos fluxes, with broader and higher-energy peaks. The reason lies in the larger amount of higher energy photons, which in turn enhance muon pair production, as seen in Fig.~\ref{fig:intRatio_XrCorona}. Higher temperatures of the X-ray background result in barely broader neutrino energy distribution, in particular at lower energies. For $\alpha = 1.5$ and $\alpha = 2$, the peaks of the energy spectra corresponding to different $T_{X}$ are roughly comparable. In contrast, for the highest photon spectral index, the peak related to $T_{X} = 10 \; \text{keV}$ is few times higher than that for $T_{X} = 1 \; \text{keV}$ one, apart from being located toward lower energies. 

\begin{figure}[htb!]
    \centering
    \includegraphics[width=0.49\textwidth]{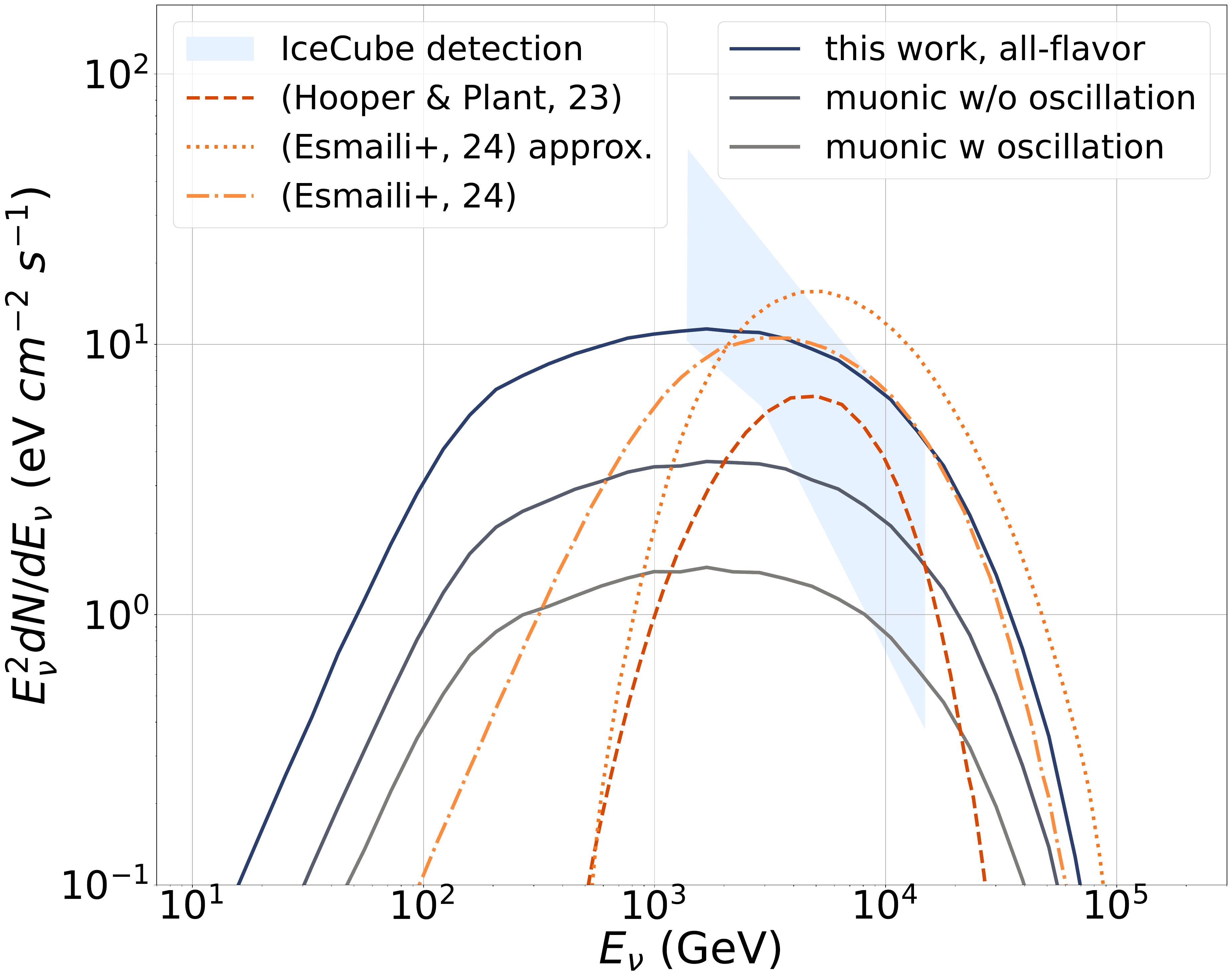}
    \caption{The results from this work are showed for the all-flavour and only-muonic spectra, this latter with and without accounting for oscillations. They are together with the results of Ref.~\cite{hooper2023leptonic} (\textit{dashed line}) and Ref.~\cite{esmaeili2024neutrinos} (\textit{dotted} and \textit{dashed-dotted lines}). The IceCube neutrino detection of NGC~1068~\cite{2022} is the \textit{shaded area}. The spectral index is fixed to $\alpha = 1.5$, while the characteristic temperature of the X-ray field is set to $1 \; \text{keV}$.}
    \label{fig:spectraAGNnuComp}
\end{figure}

Normalising the fluxes above $1 \; \text{TeV}$ to a total injected luminosity of $1.2 \times 10^{43} \; \text{erg/s}$, as in the case of $T_{X} = 1 \; \text{keV}$ in Ref.~\cite{hooper2023leptonic}, we present our benchmark energy spectra in Fig.~\ref{fig:spectraAGNnuComp}. It is compared to the mentioned previous computations. The photon spectral index  and the cutoff energy are set, respectively, to $\alpha = 1.5$ to $E_\text{max} = 20 \; \text{TeV}$. The \textit{dashed} and \textit{dotted lines} are previous computations, respectively from Refs.~\cite{hooper2023leptonic} and~\cite{esmaeili2024neutrinos}, that, like in this work, include only electron and muon pair production processes. The \textit{dashed-dotted} energy spectra refers to a computation involving the other EM processes, double and triplet pair productions among others~\cite{esmaeili2024neutrinos}. In the same Fig.~\ref{fig:spectraAGNnuComp}, we show the $\nu_{\mu} + \bar{\nu}_{\mu}$ IceCube detection of the neutrinos associated to NGC~1068~\cite{2022}.

\begin{figure}[htb!]
    \centering
    \includegraphics[width=0.55\textwidth]{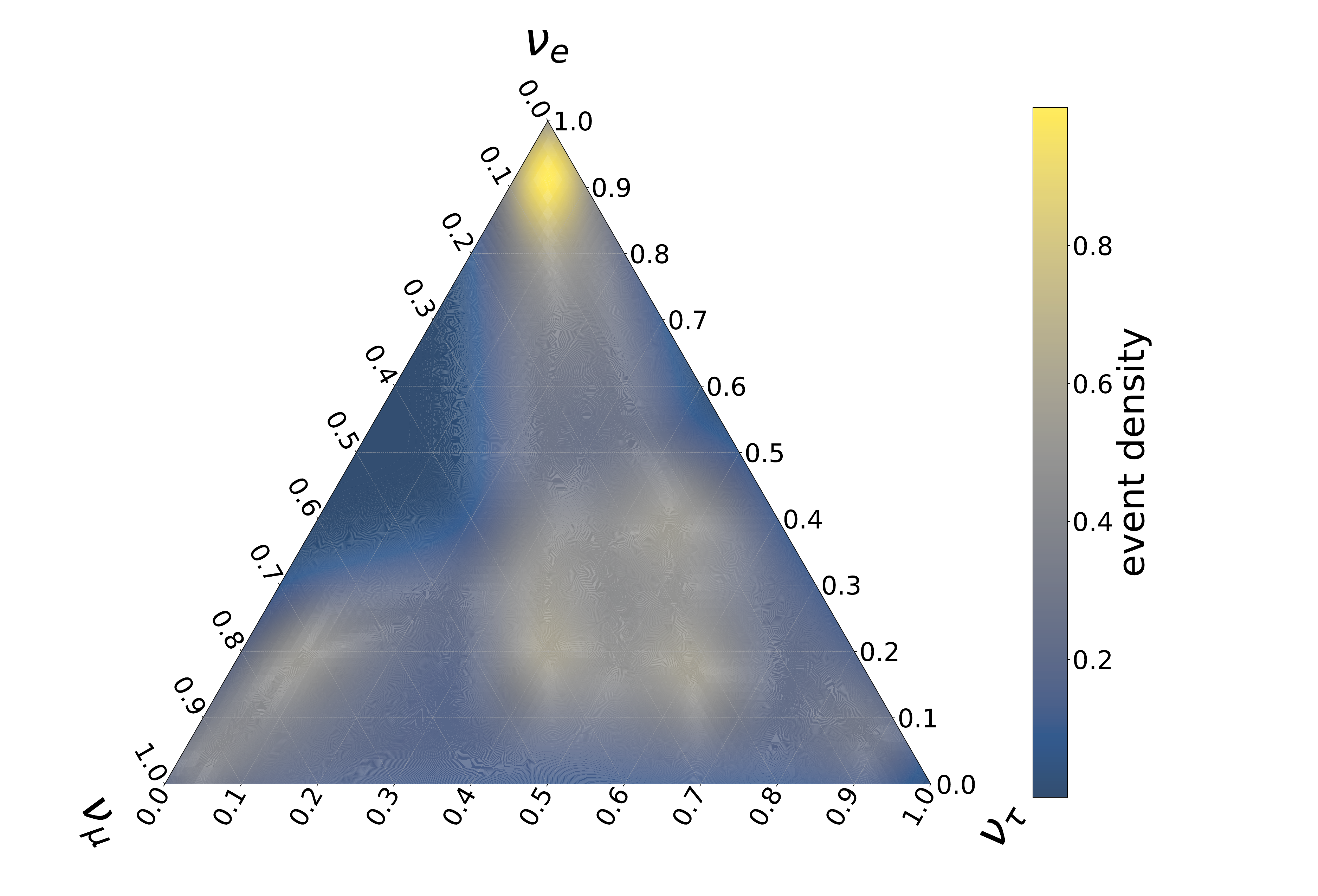}
    \caption{Flavour triangle of the \textit{leptonically-produced} neutrinos from NGC~1068, after accounting for oscillations.}
    \label{fig:flavPlotNGC}
\end{figure}

If we account for exact oscillation probabilities (see~\ref{app_propProb}) the expected flavour composition of neutrinos produced via muon pair production in the corona of NGC~1068 is shown in Fig.~\ref{fig:flavPlotNGC}. The actual flavour distribution could be crucial in determining the processes behind the neutrino production, i.e. if produced from the decay chain of charged pions or  just from muons. The former prompt composition is a signature of hadronic interactions, while the latter of leptonic production of neutrinos. In this work, we neglect any possible modified oscillation probabilities due to the surrounding AGN environment~\cite{dev2023flavor}.

\section{Discussion}\label{secDisc} 

In the previous two sections, we presented neutrinos produced during the cosmological propagation of highly-energetic gamma rays and by gamma--X-ray interactions within the cores of AGNs.

\paragraph*{Production during gamma-ray propagation} In a similar vein to Ref.~\cite{esmaeili2022ultrahigh}, we characterise the properties of neutrinos produced in  EM cascades generated by \textit{fictitious} high-$z$ monochromatic sources, as detailed in Sec.~\ref{secMonoSources}. Unlike Ref.~\cite{esmaeili2022ultrahigh}, which includes neutrinos with energies down to $1 \; \text{PeV}$ in the analysis, our simulation energy threshold is set to $10 \; \text{PeV}$. This contributes to the lower amount of energy channelled into neutrinos in the present work, as shown in Fig.~\ref{fig:enFrac_inj}, compared to previous results (see Fig.~3 of Ref.~\cite{esmaeili2022ultrahigh}). 

The main innovation of this work resides in the introduction of the CRB and EBL backgrounds. Although the interaction rates are dominated by resonances of interactions with the CMB (see Fig.~\ref{fig:intRatesCosmo}), the EBL and CRB provide significant additional targets for gamma rays and electrons at energies far from the CMB interaction rate peaks. This is particularly relevant for low center-of-mass energy EM~processes, namely electron pair production and inverse Compton scattering, as well as for their higher order counterparts (double and triplet pair production). For instance, taking into account gamma rays (or electrons) with energies $> 10^{18} \; \text{eV}$, viz. at which muon pair production on the CMB becomes relevant, the rates of Fig.~\ref{fig:intRatesCosmo} for production of heavy leptons or mesons remain lower than electron (triplet) pair production on CRB (EBL). They result in even lower energies being channelled into the observed neutrinos compared to the previous computations~\cite{esmaeili2022ultrahigh, esmaeili2024neutrinos}. It is important to note that the high-redshift interaction rates involving EBL and CRB should be interpreted with caution due to the larger model uncertainties.

We also take into account tauon pair production. Despite its low interaction rate with cosmological background photons, well below the (inverse) Hubble time, it amounts for a non-negligible fraction ($\gtrsim 1 \%$) of the detected neutrino energy for sources with $E_{\gamma, \text{inj}} > 10^{22} \; \text{eV}$. Nevertheless, the dominant contribution to the neutrino flux is due to muons generated in the cascades. This is expected due to their lower energy thresholds and the higher interaction rates with respect to other channels. This trend is reflected in the neutrino flavour composition before oscillations, approximately~$(0.5:0.5:0)$ with a tiny, few percent, excess of muonic neutrinos. This small asymmetry is attributed to the presence of charged pions in the cascades, from whose decay a~$(0.33:0.67:0)$ flavour distribution is expected. If oscillations are properly accounted for, the flavour composition becomes more equally redistributed under the assumptions of Ref.~\cite{anchordoqui2014cosmic}, as discussed in Sec.~\ref{secMonoSources}. 

In the energy spectra of Fig.~\ref{fig:nuSED}, the stable -- or in a few cases increasing -- profiles beyond the spectral peaks of the most energetic sources can be interpreted together with the observables reported in Sec.~\ref{secMonoSources}. These features may be attributed to secondary gamma rays and electrons that remain sufficiently energetic to produce muons or heavier particles through interactions with background photons. As shown in Fig.~\ref{fig:LSdist}, neutrinos from sources with $E \gtrsim 10^{23} \; \text{eV}$ tend to be produced closer to the observer. Moreover, the higher production of tauons in these cascades significantly contribute to the production of the highest-energy neutrinos, leading to a rise in the spectra beyond $E \gtrsim 10^{21} \; \text{eV}$. 

The profiles of the energy spectra from the farthest monochromatic sources of $10^{20}$ and $10^{21}$ in Fig.~\ref{fig:nuSED} are analogous to the ones obtained in the previous computation (see Fig.~3 of Ref.~\cite{esmaeili2024neutrinos}), corroborating our results.

The most energetic neutrinos fluxes, potentially distorted during their propagation by relic particle backgrounds (see~\ref{app_nuAbs}), could be investigated through lunar neutrino radio telescopes, as, for instance, the NuMoon experiment~\cite{Scholten_2006, scholten2009first} and the planned project for the ULW radio telescope~\cite{chen2023detection}. At slightly lower energy bands, the IceCube-Gen2~\cite{aartsen2019neutrino, aartsen2021icecube} and GRAND~\cite{alvarez2020giant} projects will improve the detection sensitivities of the current neutrino observatories. 

For the extreme cosmological EM cascades investigated, secondary particles produced from decays are strongly boosted forward, parallel to the parent's direction of motion. Therefore, angular corrections can be neglected, as they do not affect significantly the neutrino observables. However, extragalactic magnetic fields could impact on neutrino arrival directions. The magnitude and coherence length of these fields determines the ``deflection'' of the neutrino momentum vector with respect to the primary particle's trajectory, in a similar fashion to what happens with gamma-ray--induced EM cascades (see, for instance, Ref.~\cite{Alves_Batista_2021}). If the magnetic field coherence length is smaller than the inverse mean free path of Fig.~\ref{fig:intRatesCosmo}, only the neutrinos produced from secondary interaction of electrons or secondary gamma rays in the cascade could propagate with a different direction with respect to the prompt gamma-ray one. In the case of magnetic field sizes comparable or smaller than average distances travelled by particles before decaying, the trajectories of charged heavy leptons or hadrons could be bent, generating ``deflected'' neutrinos. Evaluating such spatial effects is beyond of the scope of this paper. 

The last remark about the magnetic field effect on cosmological EM cascade regards the electron-positron pair production from gamma-ray interactions with magnetic fields. This process is found to be negligible in this scenario, since the parameter for gamma-ray conversion in an external magnetic field\footnote{The parameter controlling electron pair production from a gamma ray of energy $E_{\gamma}$ moving through an external magnetic field of perpendicular component $B_{\text{ext}, \perp}$ is defined as $\chi=\frac{E_{\gamma}}{2m_{\text{e}}c^2}\frac{B_{\text{ext}, \perp}}{B_{\text{crit}}}$. The critical QED magnetic field intensity $B_{\text{crit}}$ has a magnitude of $\sim 10^{13} \; \text{G}$~\cite{erber1966high}.} is $\ll 1$ for the typical values of the extragalactic magnetic field, even for gamma rays beyond $10^{22} \; \text{eV}$. 

\paragraph*{Production from gamma rays inside AGNs} The leptonic production of neutrinos in AGNs (Sec.~\ref{SecNuFromAGN}) suggests a dependence on both the prompt gamma-ray and on the X-ray spectra, which depends on the temperature, as seen in the resulting neutrino energy spectra shown Fig.~\ref{fig:spectraAGNnu}. 

In the benchmark scenario of Fig.~\ref{fig:spectraAGNnuComp}, the resulting neutrino spectra range from approximately $\sim 10-20 \; \text{GeV}$ to $\sim 70 - 80 \; \text{TeV}$. The highest-energy part of the all-flavour energy spectrum falls within the IceCube measurement uncertainties, represented by the \textit{shaded area} of Fig.~\ref{fig:spectraAGNnuComp}. In the case of the muonic-only energy spectrum without accounting for flavour oscillation, the predicted flux overlaps with the highest-energy IceCube detection. However, when oscillations are considered and randomly picked from the probabilities of Fig.~\ref{fig:flavPlotNGC}, the muonic neutrino flux is further reduced. 

The broader peaks arising our simulations are explained by the fact that the only interactions at play are electron and muon pair productions. Double pair production, for instance, would lower the amount of muons produced, while charged pion or tauon pair productions are shown to be negligible in this picture (see Fig.~\ref{fig:intRatio_XrCorona}). The higher production of lower-energy neutrinos is also attributed to the Monte Carlo generation of neutrinos from the muon decay, which covers a broad energy range, i.e. two orders of magnitudes below the decaying particle (see~\ref{appDecSpec}). Likewise to previous studies, the normalisation is set for gamma rays beyond 1~TeV, so it is not affected by the lower energy gamma rays. Moreover, the simulation energy threshold is set to $10 \; \text{GeV}$. This latter is lower than the one employed in Ref.~\cite{esmaeili2024neutrinos}. 

A final remark regarding the application of this leptonic model to the observations of NGC~1068 should be provided to give proper context. According to Ref.~\cite{das2024revealing}, a fully leptonic production of neutrinos would overshoot the X-ray and/or gamma-ray detections for the emission size and magnetic energy considered. Such a scenario would also require a very large gamma-ray luminosity. It would make the leptonic production still viable, even though not suitable to consistently account for the entire AGN's multimessenger spectra. This is why hadronic production models of neutrinos and their electromagnetic counterparts are usually favoured, particularly in the case of NGC~1068. 

\FloatBarrier


\section{Conclusions \& prospects}\label{secConcl} 
In this proof-of-concept study, we introduce newly integrated tools in CRPropa that enable investigations of neutrino production in EM cascades across a variety of extreme astrophysical contexts. As pointed out in previous works~\cite{Wang_2017, esmaeili2022ultrahigh}, the production neutrinos from purely leptonic processes occurring in EM cascades can alter expected correlation between gamma-ray and neutrino fluxes from hadronic interactions. This is the case for the diffuse emission~\cite{berezinsky1975cosmic, halzen2019multimessenger}, \textit{cosmogenic} fluxes~\cite{kotera2010cosmogenic, batista2019cosmogenic} or at sources~\cite{razzaque2010high, huang2022neutrino, gagliardini2024neutrino, fang2024neutrinos}. Moreover, some neutrino detections could also be fully explained considering solely leptonic scenarios~\cite{hooper2023leptonic}, without invoking high-energy protons or other nuclei, e.g. in the case of NGC~1068 neutrino detections~\cite{padovani2024high}. Similarly, these high center-of-mass energy leptonic processes might play a role in the highest-energy gamma-ray and neutrino fluxes at Earth -- or their corresponding contributions to the diffuse backgrounds --, particularly in scenarios involving super heavy dark matter decay or cosmic string interactions. These contributions could potentially impact existing indirect constraints on such models (e.g.~\cite{lunardini2012cosmic, das2023revisiting}). Quantifying the relevance of heavy leptons or hadron production in EM cascades in these various contexts will be subject of follow-up studies. Such future investigations are planned to be performed within the standard CRPropa framework -- yet containing the tools to simulate different emission models and environmental properties -- complemented with the original plug-ins presented in this work.

We present robust and extensible computational tools, suitable in the versatile CRPropa framework, with the purpose of further extending and generalising their capabilities. Future developments will include the implementation of additional channels, such as $q\bar{q}$ and charged kaon pair production, as well as other specific processes that could become relevant in particular astrophysical scenarios.

The simulations of cosmological monochromatic sources serve as a diagnostic tool for the newly developed CRPropa plug-ins. They provide a controlled environment to evaluate the resulting neutrino fluxes from peaked, highly-energetic gamma-ray emissions, reaching energies above $10^{23} \; \text{eV}$. Throughout this work, we emphasised the critical importance of extragalactic environmental properties, in particular the full spectrum of cosmological background photons, in determining multi-messenger signatures. 

Furthermore, we explored the leptonic production of neutrinos in the environment of AGNs, comparing our results with the neutrino detections from NGC~1068 and previous calculations. Yet, only an in-depth analysis of this scenario -- comprehensive of the electromagnetic counterpart -- would really allow us to capture the complexity of AGN physics.

In the long run, we hope this work contributes to paving the way for fruitful synergies among state-of-the-art simulation codes designed for various purposes, in particular event generators used in particle physics such as PYTHIA, and multi-messenger frameworks like CRPropa. This would lead to more sophisticated studies, including, for example, QCD effects or beyond standard model interactions in astrophysical contexts, ultimately improving predictions for current and future neutrino and gamma-ray observatories, fully exploiting the power of the multi-messenger approach. 

\section*{Acknowledgements}

GDM's work is supported by \textit{FPI Severo Ochoa} PRE2022-101820 grant. GDM also thanks the Institut d'Astrophisique de Paris, where part of this work was carried out, for the hospitality. This work was funded by the ``la Caixa'' Foundation (ID 100010434) and the European Union's Horizon~2020 research and innovation program under the Marie Skłodowska-Curie grant agreement No~847648, fellowship code LCF/BQ/PI21/11830030. The work of GDM and MASC was also supported by the grants PID2021-125331NB-I00 and CEX2020-001007-S, both funded by MCIN/AEI/10.13039/501100011033 and by ``ERDF A way of making Europe''. The authors also acknowledge the MultiDark Network, ref. RED2022-134411-T. RAB acknowledges the support from the Agence Nationale de la Recherche (ANR), project ANR-23-CPJ1-0103-01. 

\section*{Data availability}

The data of both cosmological and at source EM cascade simulations are available at the \href{https://projects.ift.uam-csic.es/damasco/?page_id=831}{data repository} of the DAMASCO group website. For further details regarding the data, analysis and codes used in this work, interested readers are encouraged to contact the authors.

\bibliographystyle{elsarticle-num} 
\bibliography{main}

\newpage
\appendix
\onecolumn

\section{Extreme EM processes in CRPropa}\label{appEMprocCS}
The repositories for the plug-ins implementing photon-photon production of muons, charged pions and tauons in the CRPropa code are, respectively, in \href{https://github.com/GDMarco/EMMuonPairProduction}{\texttt{EMMuonPairProduction}}, \href{https://github.com/GDMarco/EMChargedPionPairProduction}{\texttt{EMChargedPionPairProduction}} and \href{https://github.com/GDMarco/EMTauonPairProduction}{\texttt{EMTauonPairProduction}}. 
Muon pair production via interactions between high-energy electrons and soft photons is implemented in the plug-in \href{https://github.com/GDMarco/EMElectronMuonPairProduction}{\texttt{EMElectronMuonPairProduction}}.

The interaction rates shown in Fig.~\ref{fig:intRatesCosmo} of the main text are computed using the standard CRPropa procedure\footnote{The codes where the tables are computed are at the specific branch, \href{https://github.com/GDMarco/CRPropa3-data/tree/EMCascadePlugins}{\texttt{EMCascadePlugins}}, forked from the official \texttt{CRPropa3-data} repository. The tables are computed for gamma rays or electron energies up to $10^{26} \; \text{eV}$.}. The only exception is that the generation of secondary charged pions uses the inelasticities from the MUNHECA code~\cite{esmaeili2024neutrinos}. 

\begin{figure}[ht]
    \centering
    \includegraphics[width=0.49\textwidth]{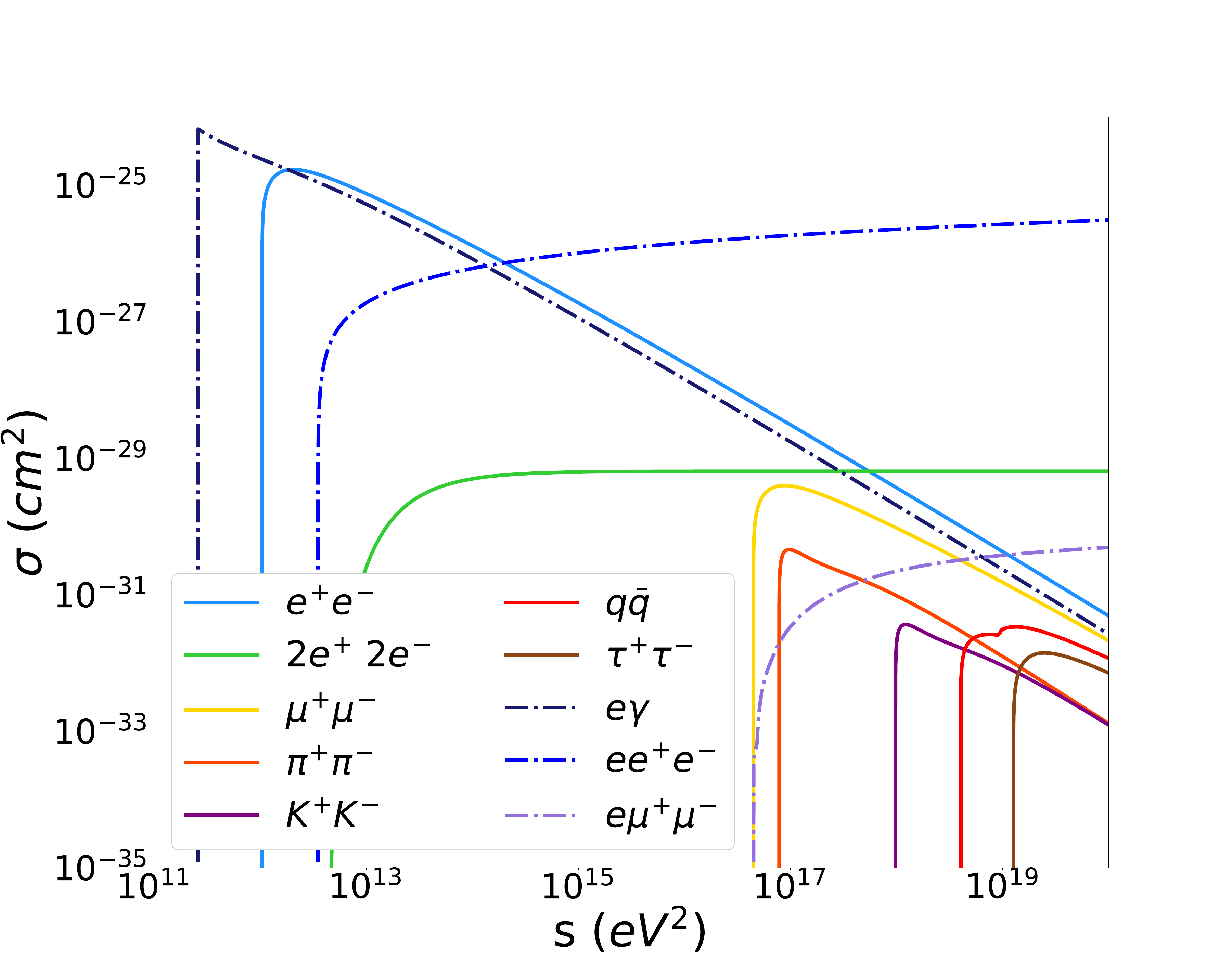}
    \caption{Cross sections of the EM processes described in Sec.~\ref{interactionSubSec}. \textit{Solid lines} are associated to photon-photon interactions, while \textit{dash-dotted} ones to the interaction between energetic electrons with low-energy photons (as in the \textit{legend}).}
    \label{fig:EMallCS}
\end{figure}

For charged pion pair production, the Born-approximated cross section is employed, neglecting any other possible strong-interaction modifications~\cite{aihara1986pion,ji1990perturbative,diehl2002handbag}. The cross-section profiles for the interactions reported in Sec.~\ref{interactionSubSec} are shown in Fig.~\ref{fig:EMallCS}. The cross section for the $q\bar{q}$ production is estimated by simulating various photon-photon collisions with PYTHIA~\cite{helenius2018photon}\footnote{In these event generations, photons are treated as \textit{unresolved}, keeping active hard QCD processes. Final states include light quark pairs as well as $b\bar{b}$ and $c\bar{c}$ production channels.}. The cross section for $K^{+}K^{-}$ production is computed with the same Born-approximation employed for charged pions. These latter two processes are not included in our work.


\section{Specificities on the decay plug-in for CRPropa}\label{appDecSpec}
In this appendix, we provide additional details about the publicly available CRPropa plug-in \href{https://github.com/GDMarco/CRPYTHIAxDecays}{\texttt{CRPYTHIAxDecays}}. It is intended to be continuously developed and generalised to include additional particles decays and processes not treated within the CRPropa framework. This plug-in evaluates the distance travelled by the unstable particles based on their lifetimes. When this happens,
the decay products are obtained using the PYTHIA event generator, and subsequently re-injected into the CRPropa simulation flow. In order to relieve the simulation from the high amount of tracked particles, the decay module developed in this work could be further optimised by employing a weighted-sampling technique, as implemented in the CRPropa EM interactions through the \textit{thinning} parameter.

\begin{figure*}[ht]
    \centering

    \begin{subfigure}{\textwidth}
        \centering
        \includegraphics[scale=0.22]{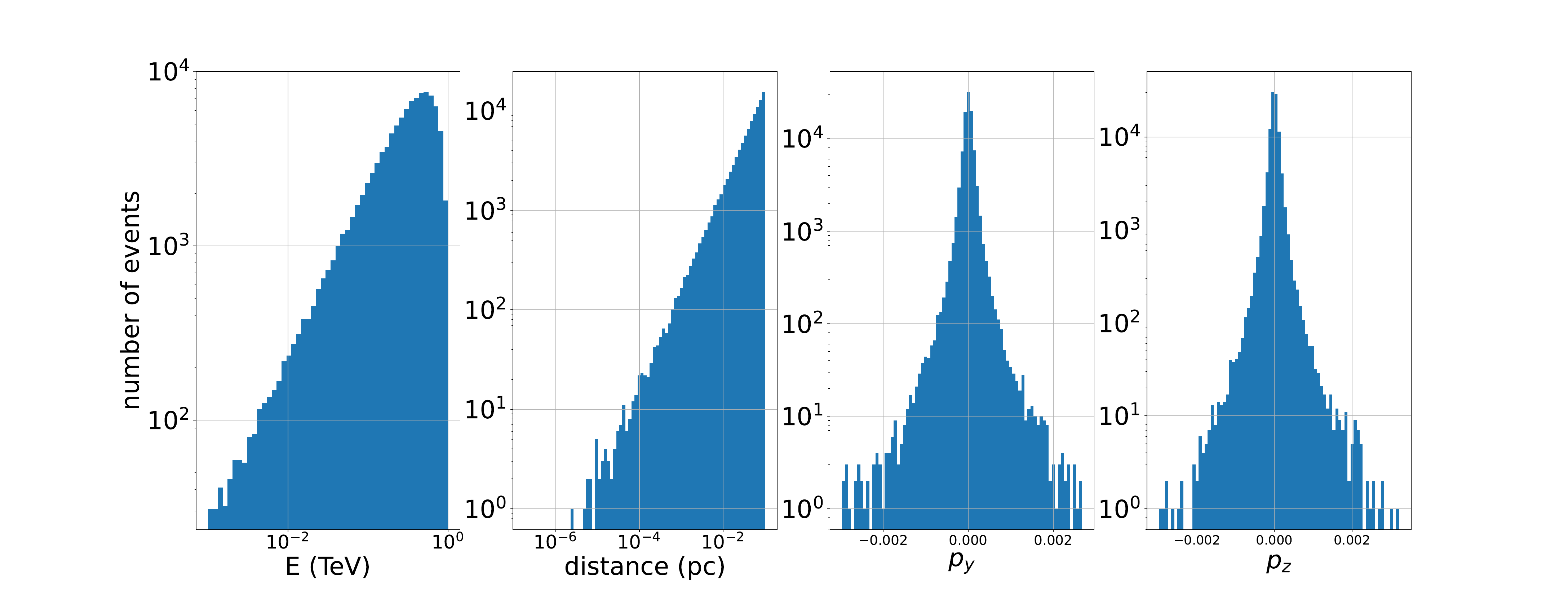}
        \caption{Properties of the muonic neutrinos, $\nu_{\mu}$, from the decay of $10^{5}$ muons ($\mu^{-}$) of $1\,\text{TeV}$.}
        \label{fig:nuMu_Mu}
    \end{subfigure}

    \vspace{0.05em} 

    \begin{subfigure}{\textwidth}
        \centering
        \includegraphics[scale=0.22]{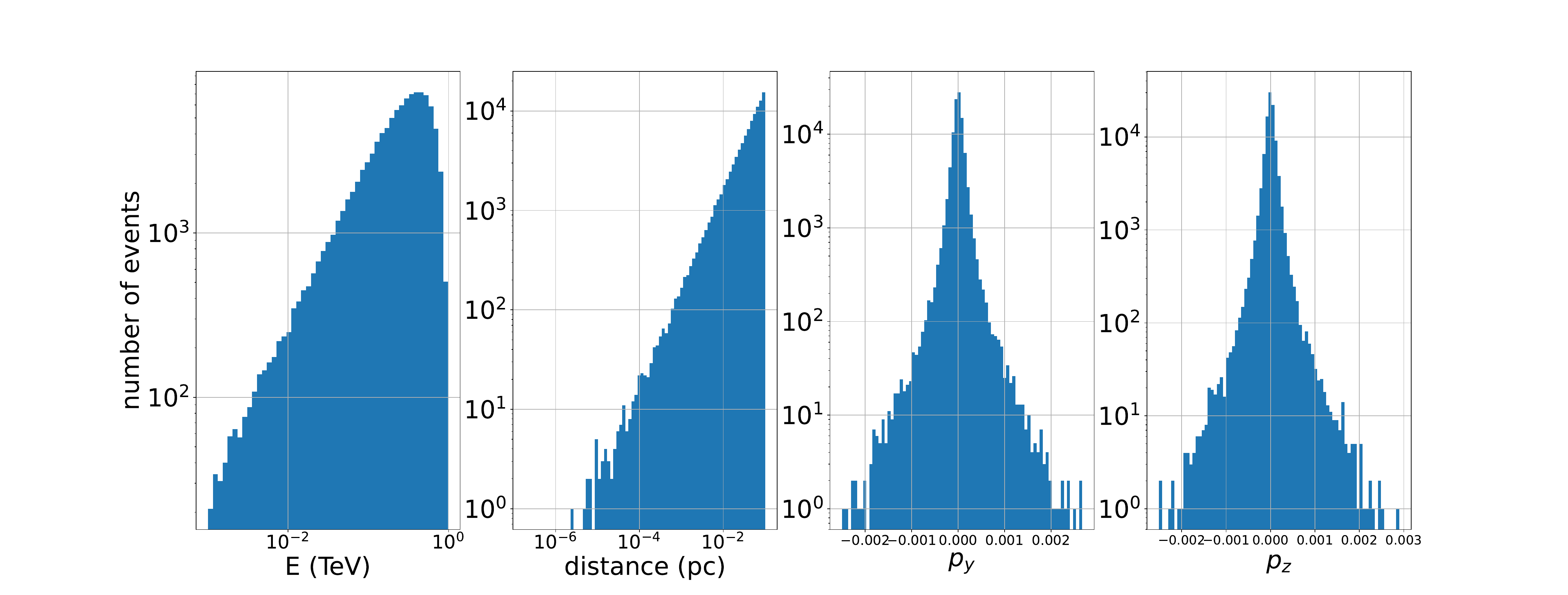}
        \caption{Properties of the electron antineutrinos, $\bar{\nu}_{e}$, from $\mu^{-}$ decay.}
        \label{fig:nuElx_Mu}
    \end{subfigure}

    \vspace{0.05em}

    \begin{subfigure}{\textwidth}
        \centering
        \includegraphics[scale=0.22]{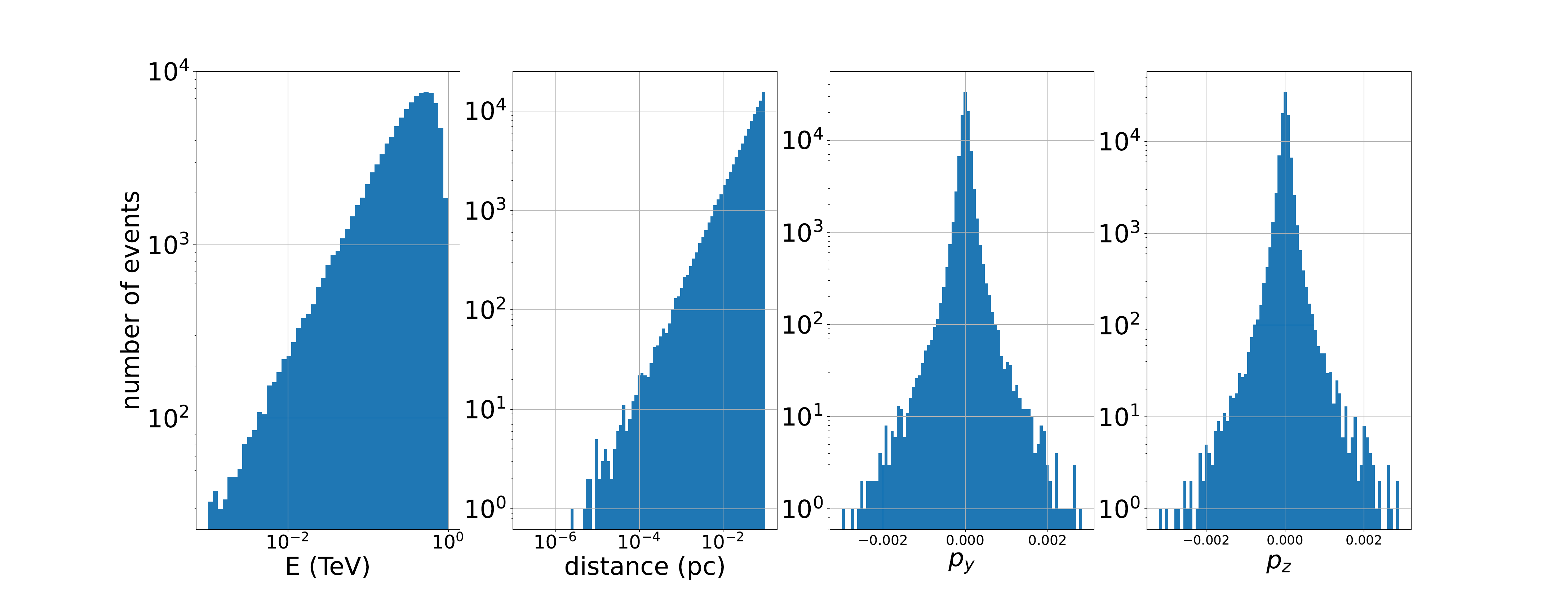}
        \caption{Properties of electrons, $e^{-}$, from $\mu^{-}$ decay.}
        \label{fig:El_Mu}
    \end{subfigure}

    \caption{Distributions of decay products from $10^5$ muons ($\mu^{-}$) of $1\,\text{TeV}$. Each row shows a different particle species with its respective kinematic properties. \textit{Left panels}: energy distribution. \textit{Center-left panels}: travelled distance. \textit{Center-right} and \textit{rightmost panels}: normalised transverse momenta of the products.}
    \label{fig:muon_decay_products}
\end{figure*}

When calling the \texttt{Decay} class, the core module of the plug-in, users can choose to generate and track neutrinos and/or other secondaries resulting from the decay. An additional flag allows users to enable or disable angular corrections in the decay kinematics. Fig.~\ref{fig:muon_decay_products} show the histograms of the properties of the decay products from $10^{5}$ muons of $1 \; \text{TeV}$.

\subsection{Tauon decay products and modes}
In CRPropa the $\gamma$-$\gamma_{\text{bkg}}$ interaction cross sections are computed by summing over all photon polarisation, effectively assuming an unpolarised beam of photons. For this reason, the polarisation-dependent $\tau$ decay modes are picked randomly by selecting the particle helicity, either longitudinal or transverse. A more accurate treatment of polarisation effects in tauon production would require the use of the appropriate PYTHIA class, \texttt{PhotonCollision:gmgm2tautau}. 
All the $\tau$ decay channels are enabled in the simulation. The dominant branching ratios correspond to the final states $\{\nu_{\tau},\pi^{0},\pi^{-}\}$, $\{\nu_{\tau},e^{-},\bar{\nu}_{e}\}$ and $\{\nu_{\tau},\mu^{-},\bar{\nu}_{\mu}\}$, with approximate branching fractions of $\sim 25\%$, $\sim 18\%$ and $\sim 17\%$, respectively.

\subsection{Decay angular corrections}
A dedicated flag allows the user to enable or disable angular corrections in the decay kinematics. In this study, we employ the \emph{collinear approximation}, wherein secondary particles are assumed to be emitted parallel to the momentum of their parent particle. For a decaying particle with Lorentz factor $\gamma_{L}$ of a decaying particle, the characteristic emission angle in the laboratory frame scales as $\sim \gamma_L^{-1}$. Thus, in the ultra-relativistic limit, decay products are effectively collimated in the forward direction.
This is an excellent approximation for highly relativistic particles, such as the ones studied here. However, it cannot be straightfowardly extended to helicity-sensitive observables or cases involving off-shell dynamics, where angular and spin correlations might be relevant.

In the muon decay histograms, shown in Figs.~\ref{fig:muon_decay_products}, the (normalised) transversal momenta with respect to the direction of the primaries, i.e. the x-axis, are shown. These distributions confirm the validity of the collinear approximation, exhibiting very small angular deflection even for TeV~muons and pions.

\section{Propagation of neutrinos}
After their production, neutrinos undergo flavour oscillations, altering their flavour composition by the time they reach Earth. Moreover, if they are energetic enough, they may interact with cosmic particle backgrounds during propagation. Both of these aspects of neutrino propagation are discussed in the following sections, and put into context within the scope of this work.

\subsection{Absorption of neutrinos}\label{app_nuAbs}
\begin{figure}[ht!]
    \centering
    \includegraphics[width=0.49\textwidth]{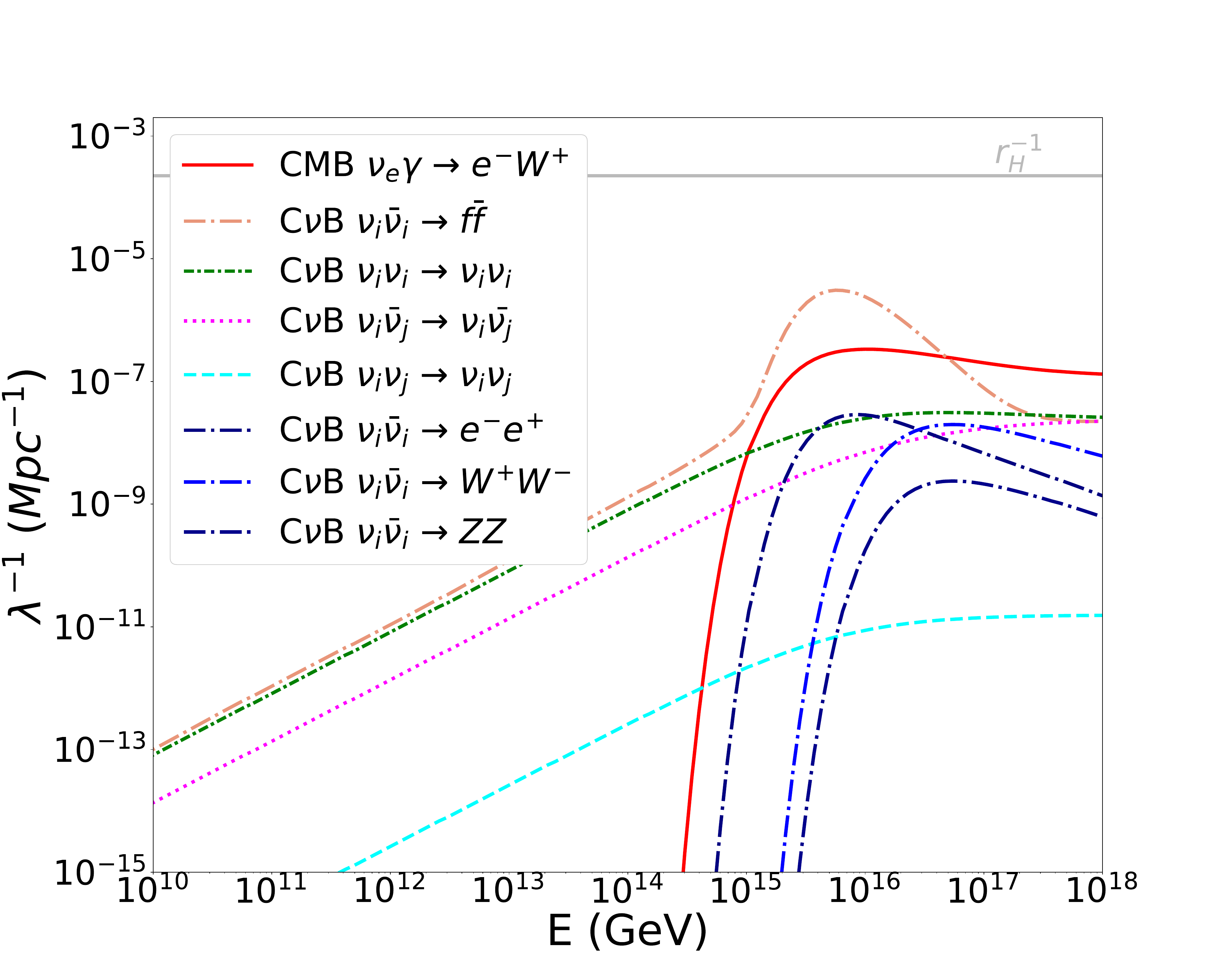}
    \caption{Inverse mean free paths for the interaction of cosmological extremely-energetic neutrinos with the CMB (\textit{red solid line}) and the C$\nu$B (in the case of massless $m_{1}$).}
    \label{fig:IMFPnu}
\end{figure}

During their cosmological propagation, the extremely-energetic neutrinos shown in Fig.~\ref{fig:nuSED} may interact with softer particle backgrounds, such as the CMB and $\text{C}\nu\text{B}$~\cite{roulet1993ultrahigh, yoshida1994propagation, seckel1998neutrino}. Trident final states from neutrino-photon interactions are possible too~\cite{zhou2020neutrino}. In this work, we neglect any potential $\nu + \gamma_\text{CMB}$ or $\nu + \nu_{\text{C}\nu\text{B}}$ interaction, even for the highest-redshift sources. This complete treatment is left for follow-up works, that would require further developments of the code. Nevertheless, in Fig.~\ref{fig:IMFPnu}, we compute the averaged interaction rates for these channels assuming a massless C$\nu$B (i.e., null first mass eigenstate $m_{1}=0$), alongside the CMB. For the other two heavier eigenstates, the resonant cross sections are expected to be narrower and shifted toward lower energies due to their modified thermal distributions~\cite{vitagliano2020grand}. 

These interactions could contribute to the energy redistribution and partial absorption of the original neutrino fluxes, especially for high-redshift sources (e.g.,~\cite{lunardini2013ultra, ruffini2016cosmic, das2024probing}). Indeed, some of the interaction secondaries might re-inject neutrinos at lower energies by decaying or by re-interacting with the same cosmological backgrounds. Moreover, neutrino fluxes might be distorted by potential overdensities in the C$\nu$B distribution~\cite{Ringwald_2004, franklin2024constraints}.

\subsection{``Propagation" of exact oscillation probabilities}\label{app_propProb}

In Sec.~\ref{secMonoSources} of the main text, the flavour composition after oscillation is briefly discussed using the approximation in Ref.~\cite{anchordoqui2014cosmic}. This calculation assumes large oscillation phases, and neglects the dependence on both neutrino energy and propagation distance.  While this approximation gives a hint of how flavour fractions are redistribute when they arrive at Earth, it neither captures the stochasticity of neutrino oscillations nor preserves energy--distance correlations. 

To illustrate the expected neutrino flavour composition accounting for exact probability calculations, we present two examples at the end of Sec.~\ref{secMonoSources}, namely Fig.~\ref{fig:cosmoTriFlav}. The ternary flavour plots are computed by explicitly propagating the neutrinos, accounting for energy- and distance-dependent oscillation probabilities, which is appropriate for cosmologically distant sources.

A neutrino of flavour $\alpha$ is produced by leptonic interactions at a redshift $z_{\text{max}}$ and eventually reach the observer at $z_{\text{obs}} = 0$. The distance covered by this neutrino is discretised in a number of steps, $N_{\text{steps}} = (z_{\text{max}} - z_{\text{obs}})  \, z_{\text{res}}^{-1}$, according to a certain redshift resolution criterion ($z_{\text{res}}$). The oscillation probability into a flavour $\beta$ is given as the average probability along this path:
\begin{equation}
    \langle p_{\alpha\beta} \rangle = \dfrac{1}{N_{\text{steps}}} \sum\limits_{i=1}^{N_{\text{steps}}}p_{i, \alpha\beta} \,,
\end{equation}
where $p_{i, \alpha\beta}\equiv p_{\alpha\beta}(E_{i}, d_{i})$ denotes the oscillation probability at the $i$-th step. The energy $E_{i}$ is redshifted according to adiabatic cosmological energy losses, and evaluated at the midpoint of each redshift interval. The interval size ($d_i$) is the comoving distance between two adjacent steps ($d_{\text{C}}(z)$, such that $d_{i}\equiv d_{\text{C}}(z_{i+1}) - d_{\text{C}}(z_{i})$. 
The exact distance- and energy-dependent probabilities are computed with the \texttt{\href{https://github.com/KM3NeT/Neurthino.jl}{Neurthino}} code~\cite{Neurthino}. The oscillation parameters are fixed to the ones reported in~Ref.~\cite{esteban2025nufit}, assuming normal mass ordering. The 
oscillation probabilities of the neutrinos coming from NGC~1068 (see Fig.~\ref{fig:flavPlotNGC} in Sec.~\ref{SecNuFromAGN} of the main text) are computed with the same tools and parameters. However, given the proximity of this source, the cosmological ``propagation'' of oscillation probabilities is not taken into account.  

\end{document}